\documentclass[utf8]{FrontiersinHarvard} 

\usepackage{url,lineno,microtype,subcaption}
\usepackage[breaklinks,colorlinks,urlcolor=blue,citecolor=blue,linkcolor=red]{hyperref}
\usepackage[onehalfspacing]{setspace}

\def\keyFont{\fontsize{8}{11}\helveticabold }
\def\firstAuthorLast{Pal {et~al.}} 
\def\Authors{Sanchita~Pal\,$^{1,*}$, Benjamin~J.~Lynch\,$^{2}$, Simon~W.~Good\,$^{1}$, Erika~Palmerio\,$^{3}$, Eleanna~Asvestari\,$^{1}$, Jens~Pomoell\,$^{1}$, Michael~L.~Stevens\,$^{4}$ and Emilia~K.~J.~Kilpua\,$^{1}$}
\def\Address{$^{1}$Department of Physics, University of Helsinki, P.O. Box 64, FI-00014 Helsinki, Finland  \\
$^{2}$ Space Sciences Laboratory, University of California--Berkeley, Berkeley, CA 94720, USA\\
$^{3}$ Predictive Science Inc., San Diego, CA 92121, USA\\
$^{4}$Smithsonian Astrophysical Observatory, Cambridge, MA 02138, USA}

\def\corrAuthor{Sanchita Pal}

\def\corrEmail{sanchita.pal@helsinki.fi}

\begin{document}
\onecolumn
\firstpage{1}

\title{Eruption and Interplanetary Evolution of a Stealthy Streamer-Blowout CME Observed by PSP at ${\sim}$0.5~AU}

\author[\firstAuthorLast ]{\Authors} 
\address{} 
\correspondance{}
\extraAuth{}
\maketitle
\begin{abstract}

Streamer-blowout coronal mass ejections (SBO-CMEs) are the dominant CME population during solar minimum. Although they are typically slow and lack clear low-coronal signatures, they can cause geomagnetic storms. With the aid of extrapolated coronal fields and remote observations of the off-limb low corona, we study the initiation of an SBO-CME preceded by consecutive CME eruptions consistent with a multi-stage sympathetic breakout scenario. From inner-heliospheric Parker Solar Probe (PSP) observations, it is evident that the SBO-CME is interacting with the heliospheric magnetic field and plasma sheet structures draped about the CME flux rope. We estimate that $18 \, \pm \, 11\%$ of the CME's azimuthal magnetic flux has been eroded through magnetic reconnection and that this erosion began after a heliospheric distance of ${\sim}0.35$ AU from the Sun was reached. This observational study has important implications for understanding the initiation of SBO-CMEs and their interaction with the heliospheric surroundings.

\tiny
 \keyFont{ \section{Keywords:} Coronal mass ejection (CME), heliosphere, Reconnection, interplanetary magnetic cloud, interplanetary magnetic field} 
\end{abstract}

\section{Introduction}
\label{sec:intro}

Coronal mass ejections (CMEs) are huge expulsions of magnetised solar plasma into interplanetary space. They were first discovered in the early 1970s \citep{tousey1973space,gosling1974mass} and initially assumed to be always associated with solar flares and/or filaments. Based on improved coronagraphic and multi-wavelength low-coronal observations, a class of CMEs emerging from streamers having signatures of flux ropes (FRs) was identified and named as `streamer-blowout' \citep[hereafter SBO;][]{sheeley1982observations, vourlidas2002analysis, vourlidas2018streamer} CMEs (SBO-CMEs). Their evacuation may take hours to days and their location mostly follow the tilt of the heliospheric current sheet \citep[HCS;][]{vourlidas2018streamer}---a boundary between open and oppositely directed heliospheric field lines \citep{smith2001heliospheric}. Having no direct association with solar active regions, flares, and/or filaments, SBO-CMEs often lack classic low-coronal signatures \citep{robbrecht2009trace, ma2010statistical, kilpua2014solar, lynch2016model} and are hence characterised as `stealth CMEs' \citep{robbrecht2009trace, howard2013stealth}. Recent high-resolution, multi-wavelength, and multi-viewpoint coronal observations have revealed weak low-coronal dynamics associated with stealth CMEs, in some cases enabling study of their formation and lift-off \citep{korreck2020source, okane2021solar, palmerio2021investigating}. Stealth CMEs may also cause significant magnetic storms at Earth that are problematic for space weather forecasting \citep{nitta2021}.

CMEs are preceded by a gradual accumulation of free magnetic energy and subsequent triggering of plasma instabilities in the energized magnetic structure. The accumulated free energy can power consecutive sympathetic eruptions in multipolar flux systems \citep{li2008solar}. In this configuration, a CME may initiate via the breakout mechanism \citep{antiochos1999}, where the breakout reconnection at an overlying, stressed null point gives rise to an increasing expansion of the energized streamer arcade that eventually triggers explosive flare reconnection below the rising sheared/twisted flux rope field lines. Under certain conditions, this multipolar topology has been shown to support consecutive eruptions from the same flux system \citep[homologous eruptions; e.g.,][]{devore2008homologous} and consecutive eruptions from adjacent flux systems \citep[sympathetic eruptions; e.g.,][]{torok2011, lynch2013sympathetic, dahlin2019model}. \cite{2012ApJ...750...12S} and \cite{2021ApJ...923...45Z} studied sympathetic filament eruptions in quadrupolar and tripolar magnetic field regions, respectively, and proposed that the magnetic implosion mechanism might be a possible link between the successive flux rope eruptions.

Although typically observed to be slow, an SBO-CME's interplanetary counterpart (ICME) may deflect and compress the ambient plasma ahead of it. This causes draping of the interplanetary magnetic field (IMF) about the ICME, where the pattern of draping depends on the overall size and shape of the ICME as well as its speed relative to the upstream plasma \citep{gosling1987field, mccomas1988interplanetary}. The draped IMF may interact with an ICME via magnetic reconnection \citep{mccomas1994magnetic, dasso2006model}, which may significantly erode its original magnetic flux and helicity \citep{dasso2006model, ruffenach2015statistical, pal2020flux, pal2021uncovering, PAL2021}.

During its fifth orbit around the Sun, Parker Solar Probe \citep[PSP;][]{fox2016solar} crossed an SBO-CME at a heliospheric distance of 0.52~AU. Along with PSP, the CME was also observed by BepiColombo and Wind \citep{mostl2021multipoint}. The CME was a stealth event that followed consecutive eruptions initiated from the Earth-facing solar disk. In this study, by multi-vantage point remote-sensing, white-light, and in-situ observations, we discuss the initiation and launch of this SBO-CME and its interaction with its surroundings in the heliosphere. 
In Section~\ref{sec:data}, we describe the satellite data sets utilised in this study.
Section~\ref{sec:overview} provides an overview of the event at its origin and in the interplanetary medium. 
In Section~\ref{sec:remote}, we present our analysis of the remote-sensing EUV and white light coronagraph data to describe the SBO-CME eruption and its coronal dynamics.
In Section~\ref{subsec:insitu}, we present our analysis of the in-situ plasma and magnetic field observations by PSP to describe the SBO-CME's heliospheric evolution.
Finally, in Section~\ref{sec:disc}, we discuss our results and present our conclusions. 

\section{Data Sets}
\label{sec:data}

To perform this study, we use remote-sensing white-light data from the inner (COR1) and outer (COR2) coronagraphs (field of views or FOVs: 1.5--4\,$R_\odot$ and 2.5--15\,$R_\odot$, respectively), part of the Sun--Earth Connection Coronal and Heliospheric Investigation \citep[SECCHI;][]{howard2008sun} onboard the Solar Terrestrial Relations Observatory \citep[STEREO;][]{kaiser2008stereo} Ahead (STEREO-A) spacecraft, and the C2 camera (FOV: 1.6--6\,$R_\odot$) part of the Large Angle and Spectrometric Coronagraph \citep[LASCO;][]{brueckner1995large} onboard the Solar and Heliospheric Observatory \citep[SOHO;][]{domingo1995soho}. We use extreme ultra-violet (EUV) imagery of the solar disk from the Extreme UltraViolet Imager (EUVI) camera onboard STEREO-A and the Atmospheric Imaging Assembly \citep[AIA;][]{lemen2012atmospheric} instrument onboard the Solar Dynamics Observatory \citep[SDO;][]{pesnell2012solar}, and photospheric radial synoptic magnetograms from the Heliospheric Magnetic Imager \citep[HMI;][]{scherrer2012helioseismic} onboard SDO. In-situ solar wind measurements are obtained from the fluxgate magnetometer, a part of the PSP's FIELDS \citep{bale2016fields} investigation, as well as the Solar Probe Cup \citep[SPC;][]{case2020solar} and the Solar Probe ANalyzers-Electron \citep[SPAN-E;][]{whittlesey2020solar} instruments, part of PSP's Solar Wind Electrons Alphas and Protons \citep[SWEAP;][]{kasper2016solar} investigation.   

\section{Event Overview} 
\label{sec:overview}

The SBO-CME first appeared over the west limb with a `classical' three-part structure, i.e., a bright front followed by a dark cavity containing a bright core \cite{illing1983possible}, in STEREO-A/COR2 images (from ${\sim}70^\circ$ east of the Sun--Earth line at ${\sim}1$ AU) around $t_{\rm start}=$ 16:30 UT on 22 June 2020. STEREO-A/COR1 and COR2 observations of a gradual swelling of the overlying streamer before the eruption and a depleted corona afterward are the characteristic signatures used to identify this eruption as a slow SBO-CME. We also identified two preceding eruptions that significantly distorted and/or disrupted the overlying coronal helmet streamer leading up to and enabling the third, slow eruption of the SBO-CME.

Figures~\ref{f1}--\ref{f3} show remote observations summarizing the source region of the stealth SBO-CME event (hereafter CME \#2) and the two preceding eruptions (hereafter CMEs \#0 and \#1). Figure~\ref{f1}(a) shows the HMI synoptic magnetogram for Carrington Rotation (CR) 2232 where the red, green, and orange vertical lines indicate the Carrington longitude of the STEREO-A, SOHO, and PSP spacecraft positions at the time of the SBO-CME, respectively. In the LASCO/C2 coronagraph, CME \#2 appeared as a faint event on the western side of the solar disc. Although the appearance of CME \#2 in the coronagraphs indicates its origin to be from the Earth-facing side of the Sun, no clear eruptive signatures were observed in SDO/AIA imagery. Therefore, we classify CME \#2 as a stealth event.  

It is apparent from EUV imagery that CME \#2 lifted off from the southern hemisphere. By modeling the coronal evolution of the CME using the Forecasting a CME’s Altered Trajectory \citep[ForeCAT;][]{kay2015global} model and confirming its results with the output of the graduated cylindrical shell \citep[GCS;][]{thernisien2006modeling} model applied to coronagraph data, \citet{palmerio2021predicting} found that the CME deflected towards the HCS. In the upper corona, we obtain a de-projected speed for the CME of $\approx$220~km/s through forward modeling with the GCS technique. The obtained speed is very similar to the typical values for stealth CMEs \citep{ma2010statistical} and consistent with the results in $\S$\ref{subsec:dynamics}. The CME arrived at PSP, located at 0.52 AU and ${\sim}20^\circ$ west of the Sun--Earth line, on 25 June, approximately three days after its detection in STEREO/COR2. At PSP the event featured a smoothly rotating magnetic field direction, high-intensity magnetic field, and low plasma-$\beta$ ($<1$) during most of its interval. No signature of a CME-driven interplanetary shock was found in the in-situ observations.



\section{Remote-sensing Observations and Analysis of the SBO-CME}
\label{sec:remote}

Off-limb STEREO-A/EUV, COR1, and COR2 imagery reveal that the SBO-CME eruption was part of a multi-stage, sequential (and most likely sympathetic) eruption scenario \citep[e.g.,][]{torok2011, schrijver2013, lynch2013sympathetic}. 
In this Section, we identify the coronal source regions and analyse the coronal dynamics of each of the sequential eruptions leading up to and including the SBO-CME.

\subsection{Solar Sources of the Sequential Eruptive Events}
\label{subsec:source}

\begin{figure}[!t]
 \centering
 \includegraphics[width=0.75\textwidth]{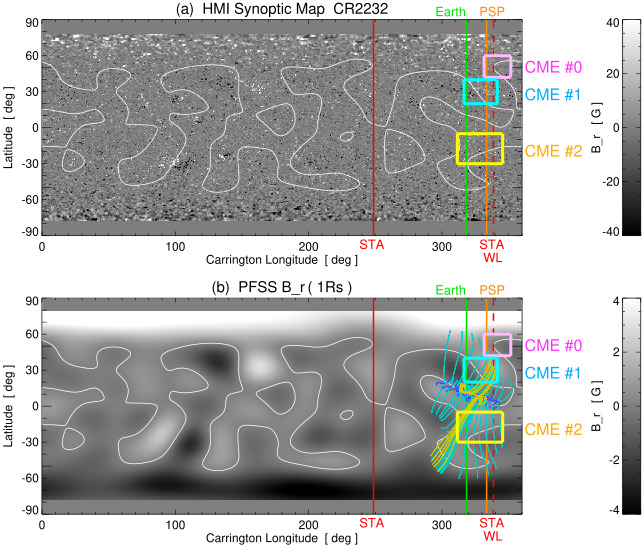}
 \caption{(a) Synoptic magnetogram of photospheric $B_r$ from SDO/HMI. The blue, green, and red vertical lines denote the Carrington longitude of STEREO-A, SOHO/Earth, and PSP on 21 June 2020, respectively. (b) Synoptic magnetogram of the PFSS $B_r$ showing the large-scale flux distributions with representative magnetic field lines of the multipolar flux system on STA's west limb (WL; red dashed vertical line). In both panels the average magnetic PILs are shown as the white contours and the three CME source regions are indicated.}
 \label{f1}
\end{figure}

\begin{figure}[!t]
 \centering
 \includegraphics[width=0.95\textwidth]{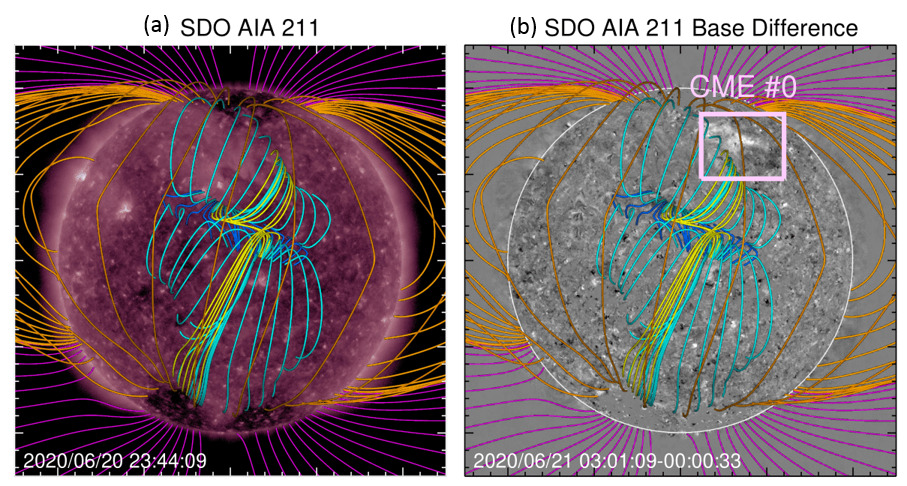}
 \caption{(a) SDO/AIA 211\AA\ emission and representative PFSS magnetic field lines illustrating the different flux systems on the Earth-facing solar disk. (b) The same PFSS field lines plotted over the base-difference 211\AA\ image. The on-disk eruption signatures (and source region) for CME \#0 (pink) are to the northwest of the multipolar flux system.}
 \label{f2}
\end{figure}

\begin{figure}[!t]
 \centering
 \includegraphics[width=0.95\textwidth]{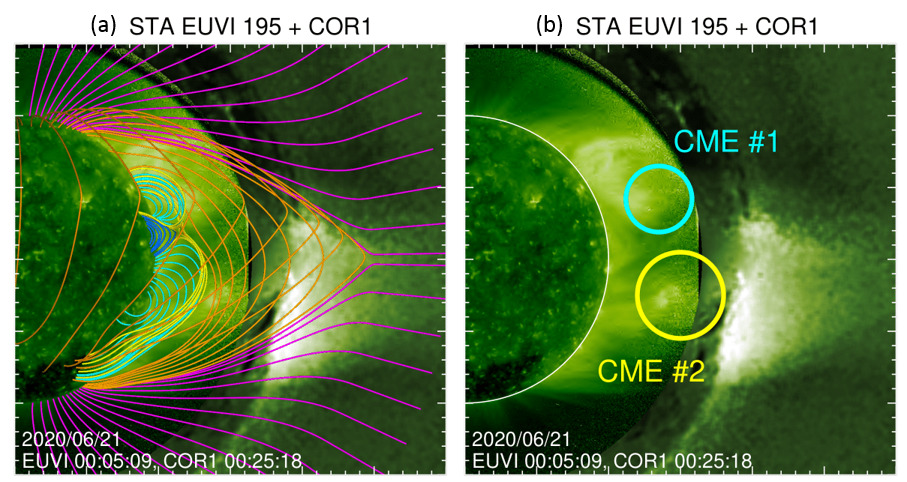}
 \caption{(a) Composite image showing the limb-enhanced STA/EUVI 195\AA\ emission together with the processed STA/COR1 white light observations. Representative PFSS magnetic field lines are also shown to identify the (approximate) positions of the different flux systems on the west limb. (b) The same composite image as panel (a) with annotation to show the pre-eruption (flux rope) progenitors for CME \#1 (cyan) and CME \#2 (yellow) at $r \sim 1.3R_\odot$ in the northern and southern side-lobe arcades, respectively. An animation of the limb-enhanced STA/EUVI data is included as Video 1 in the supplementary materials.}
 \label{f3}
\end{figure}

 The HMI synoptic map shown in Figure~\ref{f1}(a) depicts a typical solar minimum/quiet-Sun corona. Figure~\ref{f1}(b) shows a low-order potential field source surface \cite[PFSS;][]{wang1992potential} representation of the global magnetic field during CR2232. In both panels, the PFSS global polarity inversion lines (PILs) are plotted as white contours. Representative PFSS field lines are shown in Figure~\ref{f1}(b) over the synoptic map. At the longitudes of Earth and PSP (green and orange vertical lines, respectively), the latitudinal distribution of the radial field show a sequence of $+$,$-$,$+$,$-$ polarities from north to south with three main PILs on the Earth-facing disk, resulting in a multipolar configuration beneath the helmet streamer \citep[i.e. the topology required for the magnetic breakout CME initiation of][]{antiochos1999}. Figure~\ref{f2}(a) and (b) show SDO/AIA 211~{\AA} EUV image and the AIA 211~{\AA} image in base-difference to highlight the location of the on-disk signature associated with the first of the sequential eruptions, CME \#0. This region (the ${\sim}20^{\circ} \times {\sim}15^{\circ}$ area marked in pink) is located outside and to the northwest of the equatorial multipolar flux system. Representative PFSS field lines are also shown in SDO/AIA 211~{\AA} images (Figure~\ref{f2}(a) and (b)), and in a composite of STEREO-A EUVI 195~{\AA} and COR1 data (Figure~\ref{f3}(a)). Figure~\ref{f3}(a) shows the side-lobe flux systems in cyan, the central flux system in blue, the overlying flux system in orange, and open field lines in magenta. Figure~\ref{f3}(b) indicates the off-limb EUVI emission features identified as the progenitors for the second (CME \#1) and third (CME \#2) eruptions of the event sequence. The comparison of the PFSS topology in \ref{f3}(a) and the locations of the high-altitude progenitors for CMEs \#1 and \#2 in \ref{f3}(b) clearly indicate that CMEs \#1 and \#2 originate in the (energized) northern and southern side-lobes of the streamer flux system, respectively. An animation of the limb-enhanced STEREO-A EUVI 195~{\AA} imaging data is included as Video 1 in the supplementary material. In the animation, all three sequential eruptions (CMEs \#0, \#1, and \#2) are indicated using arrows, although for the animations, we recommend viewers manually sweep the slider/progress bar quickly back and forth to more easily identify the evolutionary dynamics.

\subsection{STEREO-A/COR1 Image Processing and Height--time Profiles}
\label{subsec:cor1}

To understand the eruption dynamics we analyse STEREO-A/COR1 and COR2 data and fit the height–time points calculated from the COR2 J-maps. The J-maps are time-elongation maps created following the method explained in \citet{sheeley1999}. We utilise processed COR1 data available at STEREO Science Center which gives an image every $\Delta t \sim 5$ minutes (i.e., $t_n = n\, \Delta t$). The first step is to read in all the files ($N$) and make a \textit{minimum background} image from the data sequence.
\begin{equation}
I_{\rm bkg}(i,j) = \min \left[ \, I_{\rm COR1}(i,j,t_n) \, \right] \;\;\;\; {\rm for \;} n \in [0,N-1].
\end{equation}
We note this is essentially the same procedure that is used to construct (and remove) the F-corona background component in the standard processing of SOHO/LASCO or STEREO/SECCHI coronagraph data. Our application to the previously processed COR1 data represents the removal of a minimum K-corona background. 
Next, we loop through image sequence data and (1) subtract the $I_{\rm bkg}$ minimum, (2) multiply by a (weak) radial function $\left( r_{ij}/R_\odot \right)^{\alpha}$ with $\alpha=1.2$ and $r_{ij} = \sqrt{x_i^2 + y_j^2}$, and (3) average 9 total images per frame ($\pm4$ images on either side) to obtain a $\sim$45~min averaging window. This yields
\begin{equation}
\tilde{I}_{\rm proc}(i,j,t_n) = \frac{1}{9} \sum_{n=-4}^{+4} \, \left( \frac{r_{ij}}{R_\odot} \right)^{\alpha} \, \Big( \, I_{\rm COR1}(i,j,t_n) - I_{\rm bkg}(i,j) \, \Big) \; .
\end{equation}
In the final step, we saturate the intensity range from the resulting image to enhance the contrast. This results in a type of base-difference image sequence where the (averaged) difference $\tilde{I}_{\rm proc}$ is above the intensity minimum values over a temporal interval rather than from the intensity at an initial time. In general, base-difference processing has demonstrated an advantage in the identification of larger-scale, slowly-varying signatures of eruptive transients, such as the EUV dimmings and/or brightenings, associated with stealth or stealthy CME eruptions \citep[e.g.,][]{palmerio2021investigating, nitta2021}.

\noindent{Height--time Fit Parameters}

To fit the points in the height--time ``J-maps'' derived from the STEREO-A/COR1 and COR2 measurements, we use the \citet{sheeley1999} formulation given by
\begin{equation}
\label{eq:h}
h(t) = r_0 + 2 \, r_a \, \ln{ \left[ \cosh{\left( \frac{v_a \left( t + t_0 \right)}{2 r_a} \right)} \right] } \;
\end{equation} 
that results in an analytic expression for the velocity as a function of radial distance, 
\begin{equation}
v(r) = v_a \sqrt{ 1 - \exp{\left[ \frac{-\left(r - r_0\right)}{r_a} \right]}} \; .
\end{equation}
This functional form has four free parameters, $\{ \, r_0, \, t_0, \, r_a, \, v_a \, \}$. The initial position is given by $r_0$ at time $t_0$. The parameter $v_a$ is the asymptotic velocity (as $r \rightarrow \infty$) and $r_a$ is the radial distance that $v(r)$ achieves 80\% of its asymptotic value. 

We use the IDL function {\tt curvefit.pro} to minimize the $\chi^2$ error between the observed height--time points and the model values from Equation~(\ref{eq:h}). The weights given to each point $h(t_i)$ for the fitting procedure are uniform, i.e. $w_i = 1$ for each $i$. 
The best-fit parameter values for each profile are listed in Table~\ref{table:s1} along with the velocity $v(r)$ evaluated at $r = 20R_\odot$. The height--time points are shown in Figure~\ref{f4}(g) and \ref{f4}(h) while the corresponding velocity profiles are shown in Figure~\ref{f4}(i).

\begin{table}[!t]
\caption{Best-fit parameters for the height--time profiles of the three sequential eruptions.}
\centering
\renewcommand{\arraystretch}{1.3}
\begin{tabular}{p{2cm}|cccc|c}
\hline
Event & $\;\; t_0$ [day] $\;\;$ & $\;\; r_0$ [$R_\odot$] $\;\;$ & $\;\; r_a$ [$R_\odot$] $\;\;$ & $\;\; v_a$ [km/s] $\;\;$ & $v( r = 20R_\odot)$ [km/s] \\ 
\hline
CME \#0  &  -21.12  &  2.47  &       1.26  &   350.17  &  350.17 \\
CME \#1  &  -21.58  &  1.67  &   1321.0  &   1472.8  &  172.90 \\
CME \#2  &  -22.51  &  2.07  &     16.54  &   302.96  &  246.48 \\
\hline
\end{tabular} 
\label{table:s1}
\end{table}

\subsection{Eruption Scenario and Coronal Dynamics}
\label{subsec:dynamics}

In the following section we report the low-coronal signatures of each of the three sequential CMEs and their dynamics through the FOVs of the coronagraphs. Herein, we localize features in the EUV and coronagraph images using the position angle (PA), i.e. the counter-clockwise angle from 0$^{\circ}$ pointing up (solar north).

\begin{figure}[!t]
 \centering
 \includegraphics[width=\textwidth]{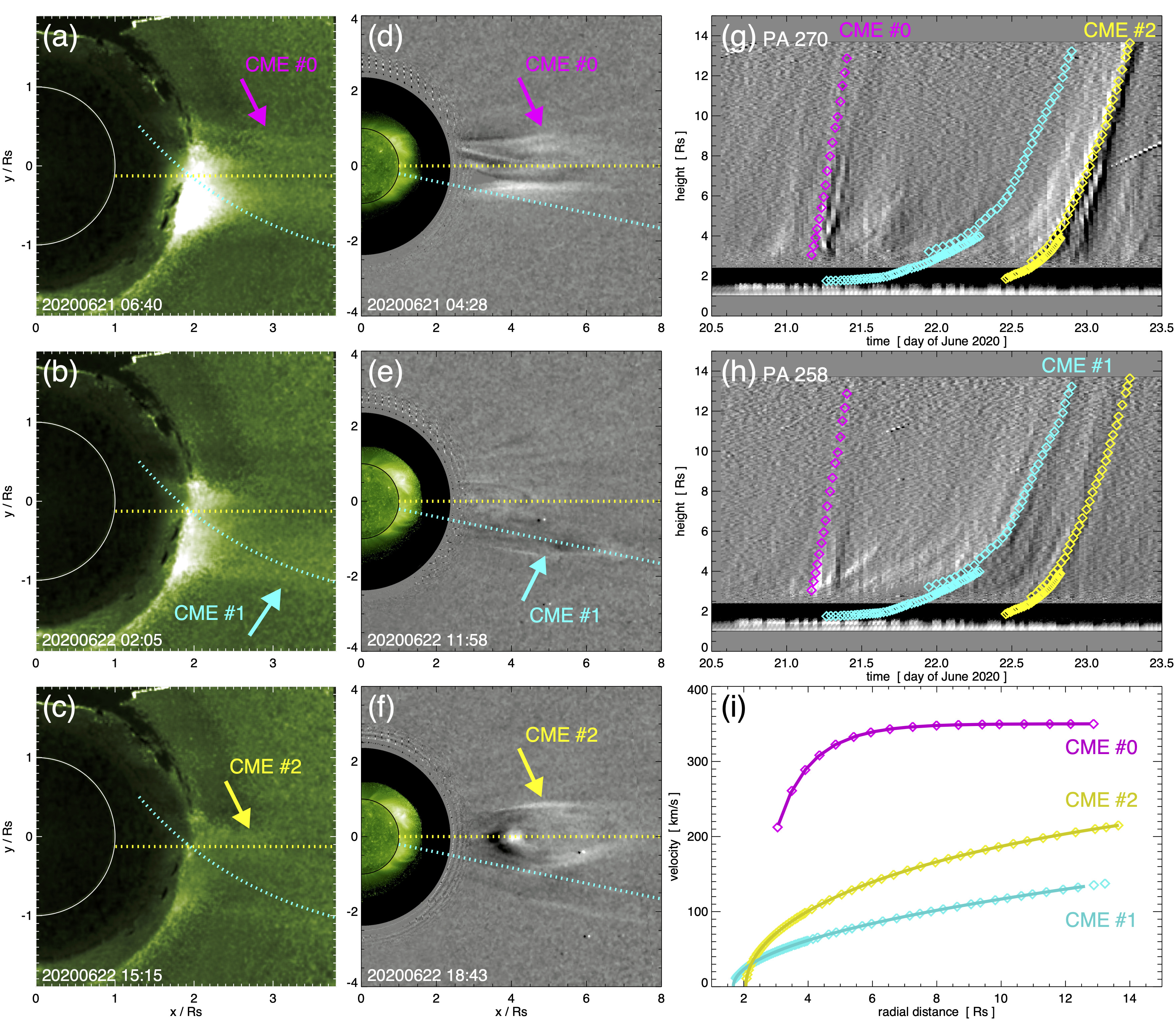}
 \caption{Coronal dynamics for each of the three sequential CMEs over the 2020 Jun 21--22 sequential/sympathetic eruption period. (a)--(c) STA/COR1 signatures of CMEs \#0, \#1, and \#2, respectively. The processed COR1 movie is available as Video 2 in the supplementary material. The cyan (yellow) dotted line shows the sampling track of CME \#1's (CME \#2's) COR1 height-time points. (d)--(f) STA/COR2 running-difference signatures of each eruption with PAs 258 (cyan) and 270 (yellow) used in the COR2 height-time points for CME \#1 and CMEs \#0 and \#2, respectively. The running difference COR2 movie is included as Video 3 in the supplementary material. (g) STA/COR2 J-map at PA 270 with the height-time points over-plotted. (h) STA/COR2 J-map at PA 258. (i) Radial evolution of the CME velocity from fits to the \citet{sheeley1999} formulation for the height-time profiles (CME \#0--magenta, \#1--cyan, \#2--yellow).}
 \label{f4}
\end{figure}

CME \#0 is the only eruption that had easily recognizable on-disk signatures in SDO/AIA data. Figure~\ref{f2}(b) shows that these signatures are largely outside of the multipolar flux system. Despite CME \#0's source region being behind the limb (from the STEREO-A view), the upper edge of the northern side-lobe arcade does brighten in the off-limb EUVI 195~{\AA} data, beginning on 21 June at 00:54~UT and followed by an apparent loop opening at $\sim$1.3\,$R_\odot$ and PA 308$^{\circ}$. The EUVI off-limb activity continues through $\sim$03:30~UT. The AIA 211~{\AA} dimming shown in Figure~\ref{f2}(b) begins on 21 June at $\sim$03:00~UT, followed immediately by a brightening that expands northward until it reaches the coronal hole boundary. The first indication of CME \#0 in the COR2 running-difference data occurs at 03:24~UT as a brightening at the northern edge of the helmet streamer. The top portion of the streamer is apparently blown out (detaches) and has a concave-up morphology in the running-difference COR2 movie (Figure~\ref{f4}(d), Video 3). By 04:54~UT, the leading edge of the CME \#0 ejecta has reached $\sim$6\,$R_\odot$ (at PAs 265--280$^{\circ}$) while the trailing edge is at $\sim$2.7\,$R_\odot$. Figure~\ref{f4} (top row) shows the CME \#0 streamer detachment eruption in STEREO-A COR1 data, COR2 running-difference data, and in the J-map height--time plot at PA 270$^{\circ}$ as panels \ref{f4}(a), \ref{f4}(d), and \ref{f4}(g), respectively. Figure~\ref{f4}(i) shows that CME \#0 quickly reaches a constant, $350$~km/s radial velocity for $r \ge 6\,R_\odot$.

The next two CMEs (\#1 and \#2) originate from the northern and southern side-lobes of the equatorial multipolar flux system. These eruptions occur much closer in time, with the upper part of the northern side-lobe EUVI enhancement in Figure~\ref{f3}(b) apparently rolling down the boundary of the central arcade structure, beginning to disconnect on 21 June at 09:24~UT (at 1.3\,$R_\odot$, PA $\sim$270$^{\circ}$), and disappearing from the EUVI field of view by 11:54~UT. Signatures of CME \#1 in the COR1 field of view follow an overall southwestern trajectory (dotted cyan line in \ref{f4}(a)--\ref{f4}(c)). A narrow, dark cavity appears at the inner edge of the COR1 occulter on 21 June at $\sim$08:00~UT (1.75\,$R_\odot$, PA 280$^{\circ}$), just after the opening of the northern flank of the helmet streamer by CME \#0 in Figure~\ref{f4}(a). The cavity expands to the southwest and an extremely faint, circular leading edge becomes visible at $\sim$16:00~UT. As this front expands, the streamer swells and deflects towards the south. This front fades with distance, becoming indistinguishable from the streamer by 23:40~UT. A faint, circular arc structure is seen at the inner boundary of COR2 (3--3.3\,$R_\odot$, PA 255--275$^{\circ}$) at 20:54~UT on 21 June and a bright edge that is wider than the helmet streamer is seen to be moving along the streamer stalk at 3.6--3.8\,$R_\odot$ by 22 June 01:54~UT. The streamer continues its southward deflection in the COR2 data from 22 June 02:00--10:00~UT. A streamer blob-like enhancement begins to form/pinch-off at 11:54~UT ($\sim$5.5\,$R_\odot$, PA 258$^{\circ}$; Figure~\ref{f4}(e)). The tail end of the blob reaches $\sim$7\,$R_\odot$ by 13:54~UT. CME \#1's height--time points along the cyan dotted lines are shown on the J-maps of \ref{f4}(g) and \ref{f4}(h). The CME \#1 is the slowest of the sequential eruptions, reaching $\sim$140~km/s by 13\,$R_\odot$.

The remainder of the remote-sensing analysis focuses on the third sequential/sympathetic eruption, CME \#2 i.e., the SBO-CME. The circular, pre-eruption EUVI enhancement labeled in Figure~\ref{f3}(b) occurs over PA 255--265$^{\circ}$ at an altitude of 1.3--1.6\,$R_\odot$ and may correspond to CME \#2's pre-eruption flux rope core/central axis. This feature shows a slow, rolling motion, beginning at 05:40~UT on 21 June (see Video 1). There are other dynamic features also seen in the southern side-lobe arcade before CME \#2's eruption. For example, a bright loop begins to contract/shrink in response to CME \#1's activity in the northern side-lobe arcade which occurs simultaneously with CME \#2's flux rope progenitor beginning to rise with a slight northwestern trajectory. This EUV structure disappears from the EUVI field of view at 17:54~UT. By 19:55~UT, there is a slight enhancement of a Y-shaped feature with the vertical segment at PA 256$^{\circ}$ and the V-shaped split at 1.27\,$R_\odot$, PA 260$^{\circ}$. In COR1, the remaining helmet streamer material diminishes in brightness between CME \#1 and the eruption of CME \#2. Another V-shaped emission structure becomes visible at 11:15~UT on 22 June, at $\sim$2\,$R_\odot$, PA 270$^{\circ}$. The COR1 V-shape shows rapid acceleration and leaves the frame by 22 June at 19:25~UT. The leading edge of CME \#2 becomes visible in COR2 data at 14:13~UT. By 14:55~UT, the circular core appears at the inner boundary (2.6--3.2\,$R_\odot$, PA 264--278$^{\circ}$). The 3-part CME structure is clearly visible in Figure~\ref{f4}(f) and its kinematics are captured in the J-maps and velocities of Figure~\ref{f4}(g)--\ref{f4}(i). CME \#2 shows a larger acceleration than CME \#1, reaching a speed of $\sim$220~km/s by 14\,$R_\odot$.

\section{In-situ Observations and Analysis of the SBO-CME}
\label{subsec:insitu}

\begin{figure}[!t]
 \centering
 \includegraphics[width=0.88\textwidth]{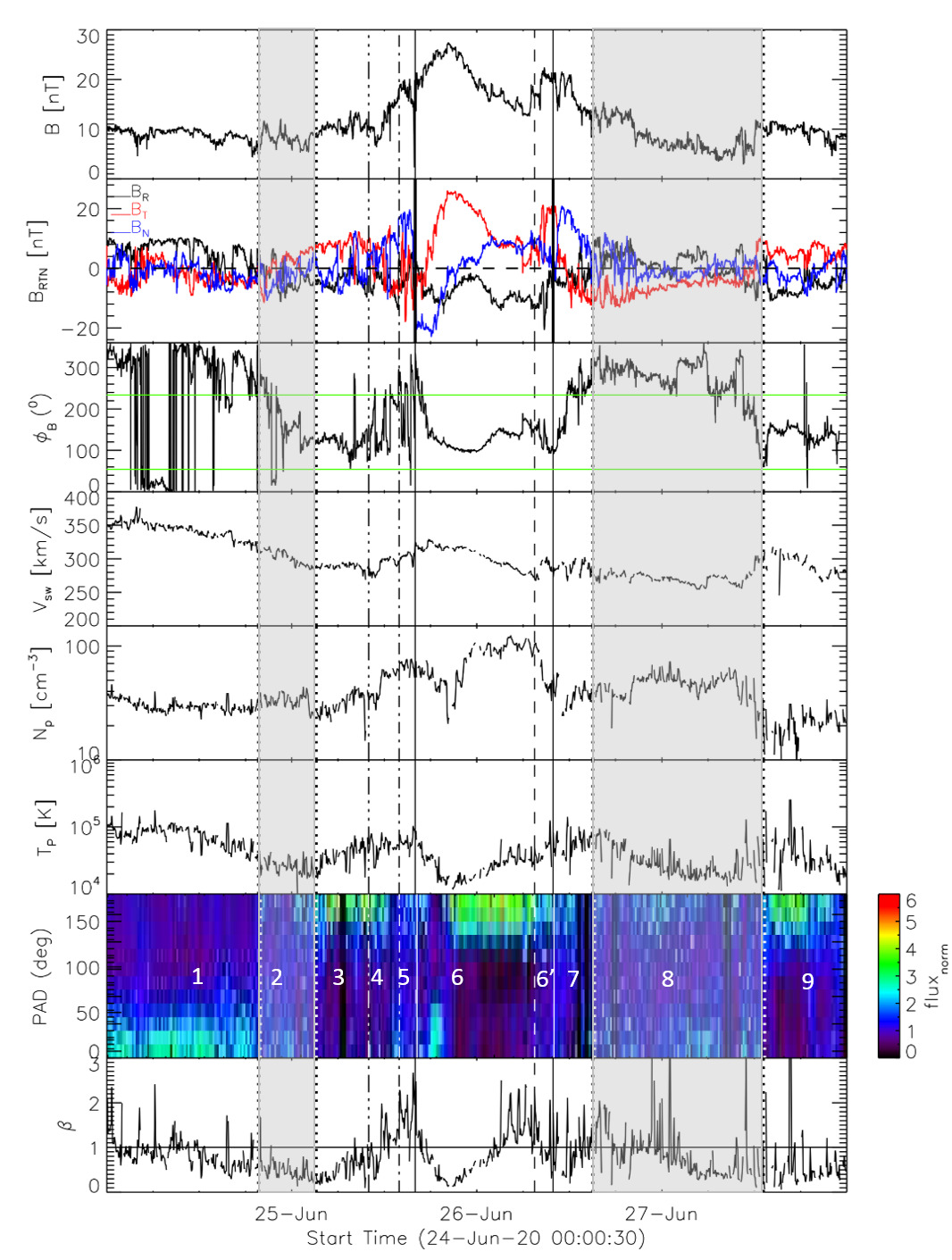}
 \caption{Solar wind magnetised plasma measurements (from top to bottom: $B$, $B_\mathrm{RTN}$, $\phi_B$, $V_\mathrm{sw}$, $N_p$, $T_p$, PAD of normalised suprathermal electron flux, plasma-$\beta$) during 24--27 June 2020. The $\phi_B$ between horizontal green lines correspond to sunward IMF. The solid black lines show $t_\mathrm{in}$ and $t_\mathrm{out}$ and the dashed vertical line indicates $t^*_\mathrm{out}$. The shaded regions correspond to HCS/HPS crossings. The annotated regions indicate the regions 1--9 described in text.  }
 \label{f5.1}
\end{figure}

Figure~\ref{f5.1} shows PSP in-situ observations of the solar wind during 24--27 June 2020. On 25 June, the PSP magnetometer recorded a steady enhancement in magnetic field magnitude $B$, the initiation of a smooth rotation in the magnetic field vector (as indicated by the $B_T$ and $B_N$ variation), and an increment in the plasma velocity $V_\mathrm{sw}$. Based on the in-situ signatures, we estimate that the leading-edge of the ICME encountered the spacecraft at $t_\mathrm{in} =$ 16:00~UT on 25 June. The coherent rotation in the magnetic field longitude angle $\phi_B$ (measured between $\mathbf{B}$ projected onto the $R$--$T$ plane and $\hat{R}$) lasted for approximately $18$ hours, until $t_\mathrm{out} =$ 09:54~UT on 26 June. The black solid vertical lines in Figure~\ref{f5.1} indicate $t_\mathrm{in}$ and $t_\mathrm{out}$. A drop in proton temperature $T_p$, a dip in plasma density $N_p$, and plasma-$\beta$ (specifically in the flux rope core) during this interval indicate the presence of a confined plasma structure, i.e., a magnetic cloud (MC). The MC was expanding as it passed PSP, the front-to-rear speed difference being  $\approx$22~km/s. Inside the MC, the electron pitch angle distribution (PAD) at 283.9--352.9~eV remains unidirectional, indicating the abundance of outward field lines at the front and inward field lines at the rear part of the MC \citep{carcaboso2020characterisation}.

\subsection{Minimum Variance Analysis and ICME Magnetic Flux Erosion}
\label{subsec:mva}

We employ nested-bootstrap minimum variance analysis \citep[MVA;][]{sonnerup1967magnetopause, kawano1995bootstrap} method and a self-similarly expanding linear force-free FR (LFF) model having cylindrical cross-section \citep{burlaga1988magnetic, lepping1990magnetic, hidalgo2000model, marubashi2007long} to the in situ observations of the magnetic profile of MC. Following \citet{ruffenach2015statistical}, we perform 1000 random data re-sampling and repeat this for seven nested time intervals separated by 10 mins within the MC. Here, each time interval starts 10 mins after and ends 10 mins before the previous time interval. Utilising these methods to MC in situ observations, we determine the MC axis orientation ($\theta_\mathrm{ax}$, $\psi_\mathrm{ax}$) in RTN coordinates.

The bootstrap method \citep{kawano1995bootstrap} with random data re-sampling helps assessing the impact of the intrinsic variability inherent in the magnetic field data on the axis determination and the nested time intervals within the MC mitigates the uncertainty involved in defining the MC boundaries on the determination of its axis. After obtaining the MC frame ($\hat{x}_\mathrm{cloud}$, $\hat{y}_\mathrm{cloud}$, $\hat{z}_\mathrm{cloud}$) using MVA and the LFF model, the magnetic field components $B_{x,\mathrm{cloud}}$, $B_{y,\mathrm{cloud}}$, $B_{z,\mathrm{cloud}}$ in the MC frame are obtained and the closest distance $p_0$ (normalised to the MC radius $R_{mc}$) between the MC center and PSP crossing path is approximated. In MVA, $p_0$ is approximated as $p_0=\langle B_{x,\mathrm{cloud}}\rangle/\langle B \rangle$ \citep{demoulin2009magnetic,ruffenach2015statistical}.

We apply the model-independent `direct method' \citep{dasso2005linking} that estimates MC flux accumulated in the azimuthal plane,  $\phi^{p,acc}$,  directly from the observed magnetic field and plasma speed profiles in MC, assuming cylindrical symmetry for the magnetic field configuration in the MC's cross-section. This method computes azimuthal flux in the outbound path (from the MC center at $t=t_c$ to its boundary at $t=t_\mathrm{out}$) as
\begin{equation}\label{eq7}
    \frac{\phi^{p}_\mathrm{dir,out}}{L_\mathrm{mc}}= \int_{t_c}^{t_\mathrm{out}} B_{y\mathrm{,cloud}}(t) \, \frac{L'_\mathrm{mc}(t)}{L_\mathrm{mc}} \, v_{x\mathrm{,cloud}}(t) \, dt \;, 
\end{equation}
where $L'_\mathrm{mc}(t)$ is the axial length of the FR, $L_\mathrm{mc}=L'_\mathrm{mc}(t_\mathrm{in})$, at the time of the spacecraft encounter, $t=t_\mathrm{in}$ \citep{dasso2005model,dasso2006model,dasso2007progressive}. To calculate $\phi^{p}_\mathrm{dir,out} \, / \, L_\mathrm{mc}$, we utilise MC axis orientation ($\theta_{ax},\psi_{ax}$) from the nested-bootstrap MVA and LFF fitting (and their mean). To perform the azimuthal flux $\phi^p_\mathrm{dir,in}$ calculation for the inbound path (when PSP approaches the MC center), the integration limits [$t_c$, $t_\mathrm{out}$] are replaced by [$t_\mathrm{in}$, $t_c$]. In Equation~\ref{eq7}, the ratio $L'_\mathrm{mc}(t) \, / \, L_\mathrm{mc}$ can be approximated as 

\begin{equation}
    \frac{L'_\mathrm{mc}(t)}{L_\mathrm{mc}} \approx 1 + \frac{(t-t_\mathrm{in}) V_\mathrm{mc}}{r_\mathrm{psp}} \; ,
\end{equation}
where $r_\mathrm{psp}$ and $V_\mathrm{mc}$ denote the Sun--PSP distance and the central speed of the MC, respectively \citep{dasso2007progressive}.

To quantify the azimuthal flux eroded due to reconnection, we calculate the inbound--outbound flux asymmetry. If reconnection occurs at the front of the MC, the reconnected flux that still remains part of the MC causes an outbound asymmetry in $\phi^{p,acc}$. Therefore, the azimuthal flux contained in the FR before the erosion is $\phi^p_\mathrm{dir,out}$ and the eroded flux is $\phi^p_\mathrm{erod}=(\, \phi^p_\mathrm{dir,out} - \, \phi^p_\mathrm{dir,in} \,)$. We determine the percentage of erosion as $\phi_{er}=( \phi^p_\mathrm{erod} \, / \, \phi^p_\mathrm{dir,out} )\,\%$. We determine a new FR boundary at $t=t^*_\mathrm{out}$ by excluding the piled up, reconnected flux at the MC rear. The region between $t^*_\mathrm{out}$ and $t_\mathrm{out}$ represents the back region of MC \citep[e.g.,][]{dasso2006model, kilpua2013relationship} that expands due to velocity difference between the solar wind velocity $V_\mathrm{sw}$ at $t_\mathrm{out}$ and $V_\mathrm{mc}$, and any additional flux that may have resulted from continued reconnection in the trailing eruptive flare/CME current sheet. 
Considering only the first factor, we estimate the time $t_\mathrm{rec}= t_{in}-\,\delta t$ after which the ongoing flux erosion starts. Here the elapsed time is given by

\begin{equation}
    \delta t = \frac{\left( t_\mathrm{out} - t^*_\mathrm{out}\right)\, V_\mathrm{sw}(t_\mathrm{out})}{V_\mathrm{mc}-V_\mathrm{sw}(t_\mathrm{out})}.
\end{equation}

\subsection{SBO-CME Adjacent Interplanetary Structures}
\label{subsec:interplanetary}

\begin{figure}[!t]
 \centering
 \includegraphics[width=0.98\textwidth]{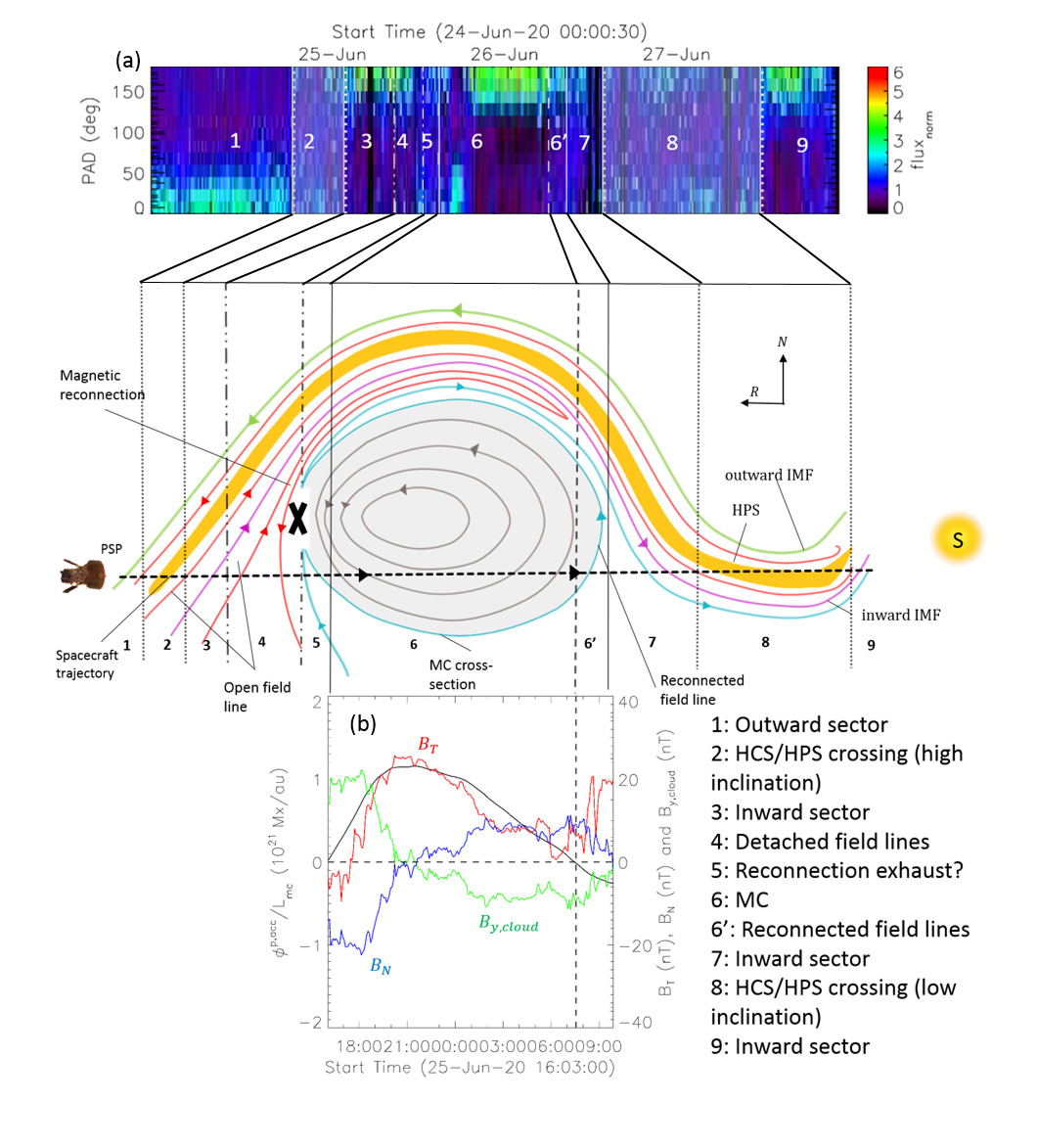}
 \caption{(a) The PAD of the normalized suprathermal electron flux during 24--27 June 2020 and a schematic (not to scale) showing the MC cross-section (grey) and adjacent interplanetary structures as observed by PSP. The regions 1--9 are indicated on the plots. The inward and outward IMFs connected to the solar origin are shown in purple and green, respectively. IMFs indicated in red and cyan correspond to open field lines disconnected from their solar origins and reconnected with MC, respectively. The HCS/HPS is shown in yellow and the PSP propagation path (in the FR frame of reference) is shown in dashed black line. (b) The time distribution of accumulated azimuthal flux ($\phi^{p,acc}/L_{mc}$), azimuthal field line ($B_{y,cloud}$) and magnetic vectors ($B_N, B_T$). the dashed line indicates $t^*_\mathrm{out}$.    }
 \label{f5.2}
\end{figure}

In Figure~\ref{f5.2}(a), we present a schematic of the MC and its adjacent interplanetary structures along with the PAD of normalised suprathermal electron flux as observed by PSP. Here and in the following discussion, the segments indicated by numbers 1--9 correspond to the annotated regions of Figure~\ref{f5.1}.

The continuous probing of solar wind preceding and following the MC indicates the presence of large-scale interplanetary structures around the MC. PSP came across an outward field line sector (1) until 19:42~UT on 24 June, when it encountered a true sector boundary (TSB; 2)---a boundary between the magnetic field lines of true opposite polarity at their solar source \citep{kahler1994determination,crooker1996solar,kahler1996topology}. It is identified by a switch in direction of suprathermal electrons from field-aligned to anti-field-aligned. The values of $\phi_B$ between the horizontal green lines (i.e., $54^{\circ}<\phi_B<234^{\circ}$) in the third panel of Figure~\ref{f5.1} correspond to the inward-directed field with respect to the Parker spiral field with a local spiral angle of $36^{\circ}$. During 2, $\phi_{B}$ displayed a ${\sim}180^\circ$ change across one of the nominal sector boundaries (green lines). Also, the suprathermal electron PAD becomes isotropic, indicating a drop in heat flux \citep{mccomas1989electron} during that time. Thus, 2 (19:42~UT 24 June -- 03:11~UT 25 June) shown by the first shaded region in Figure~\ref{f5.1}, contains the HCS crossing \citep[e.g.,][]{crooker2004heliospheric,lavraud2020heliospheric}. The coincidence of the TSB, current sheet, and elevated plasma $\beta$ at scales from minutes to few hours indicates the presence of a steady-state heliospheric plasma sheet \citep[HPS;][]{crooker2004heliospheric} encasing the HCS. The front of 2 coincides with a minor ${\sim}25$~min $\beta$ peak. This confirms the presence of the HPS in that region. Also, 2 corresponds to a high $N_p$ and low $T_p$ compared to the surroundings as well as a negative velocity gradient, which generally characterize the isotropic PAD region \citep{mccomas1989electron}. Magnetic reconnection across the HCS above a coronal streamer may generate detached magnetic structure and isotropize the suprathermal electron PAD \citep{mccomas1989electron,gosling2005magnetic,chollet2010reconnection,crooker2003suprathermal,carcaboso2020characterisation}. Due to the lack of good-quality, high-resolution data, we are unable to locate reconnection exhaust signatures \citep{gosling2005direct,phan2020parker} at the front and rear boundary of 2. We employ MVA to the magnetic field data over 2 and estimate the absolute latitude angle $\theta_n$ of the direction normal to the local current sheet \citep{lepping1996summary,lepping1996large}. We obtain $\theta_n$ as 49$^\circ$, indicating that the HCS surface was inclined to the spacecraft orbital plane. This was likely resulting from the IMF draping about the MC, which may lead to MC erosion via magnetic reconnection.

After 2, there exists an inward field line sector (3) followed by a short interval of isotropic PAD region (4). Just before the MC front boundary, during 13:55--16:00~UT on 25 June (5), we locate a region  with an accelerated $V_{sw}$, higher $N_p$, higher $T_p$, and reduced $B$ than the surrounding solar wind. These indicate a possibility of the presence of reconnection exhaust \citep{gosling2005direct}. However, due to the insufficient resolution of plasma data this cannot be confirmed. The dash-dotted line in Figure~\ref{f5.1} indicates the start time of 5. After the passage of the MC (6 and 6'), there exists an inward IMF region (7) followed by region 8 (15:03~UT 26 June -- 13:11~UT 27 June; second shaded region in Figure~\ref{f5.1}) with enhanced $N_p$, reduced $T_p$ than the surroundings, and isotropic distribution of solar wind electrons, where $\phi_B$ hovered near the boundary between the outward and inward sectors until it made a definite turn towards the inward sector at 10:50~UT on 27 June accompanied by a sharp peak in $\beta$. Taken together, these signatures suggest that once again PSP encountered the HCS and HPS. During 8, $\theta_n$ is found to be 61$^\circ$, indicating a low inclination of region 8 to the PSP orbital plane. After 8, there exists a region (9) of inward IMF.

\subsection{ICME Interaction with the Draped Interplanetary Magnetic Field}
\label{subsec:icmerxn}

To determine whether the IMF draping evident from the in-situ observations resulted in erosion of MC flux via reconnection, we further analyse the MC's magnetic flux profile. We only focus on its azimuthal flux, because erosion mainly affects the MC's outer part where azimuthal flux dominates. After obtaining the MC's axis orientation using nested-bootstrap MVA and LFF fitting, we derive the MC's accumulated azimuthal flux $\phi^p_{\rm dir, out} / L_{\rm mc}$, its eroded azimuthal flux $\phi^p_{\rm erod}$, percentage of erosion $\phi_{\rm er}$ and the time $t_\mathrm{rec}$ after which the ongoing flux erosion might have started. Both the MVA and LFF fitting techniques yield a right-handed ICME flux rope, with $p_0$  being approximately half and the flux rope axis almost parallel to the spacecraft orbital plane and perpendicular to the Sun--PSP line ($\hat{r}$).

In Table \ref{table:s2}, we summarise the results obtained from the in situ observation analysis of the MC. We provide nested-bootstrap MVA and LFF fitting results and MC's azimuthal flux balance-related analysis. Utilising the ICME's Sun-to-PSP transition speed $V_{tr}\sim300$~km/s and $t_\mathrm{rec}$, we obtain the heliospheric distance $r_\mathrm{rec} = V_{tr}\, ( \, t_{rec}-\, t_{\rm start} \,)$ after which the ongoing flux erosion might have started. We provide the result in Table \ref{table:s2}. Note that erosion of the original flux rope might have started closer to the Sun, with the corresponding back region having lost identifiable CME properties (and having become fully detached from the CME) by the time of observation at PSP. In Figure~\ref{f5.1}, $t^*_\mathrm{out}$ is indicated using a dashed vertical line. Here, the region 6' between $t^*_\mathrm{out}$ and $t_\mathrm{out}$ corresponds to the MC's back region. We notice that some of the MC characteristics such as high $B$ with low variance, coherent rotation in the field vector and a low $T_p$ continued well after $t^*_\mathrm{out}$. Figure~\ref{f5.2}(b) shows the $B_T$ (red), $B_N$ (blue), azimuthal field component $B_{y,\mathrm{cloud}}$ (green), and accumulated azimuthal flux $\phi^{p,acc} \, / \, L_\mathrm{mc}$ (black) of the MC as functions of time. Here, The dashed vertical line indicates $t^*_\mathrm{out}$. 
 
\begin {table}
\caption{Summary of results obtained from the ICME in-situ flux analysis at PSP.\\[0.40cm]}
\centering
\renewcommand{\arraystretch}{1.6}
\begin{tabular}{ l|c|c|c }
\hline
\hline
Quantity & \multicolumn{3}{c}{In-situ Reconstruction Method} \\
 \hline
 & {LFF} & {MVA}$^{\; a}$ & mean( {LFF}, {MVA} ) \\
\hline
\hline
Time interval of MC & \multicolumn{3}{c}{16:00 UT 25 June -- 09:54 UT 26 June} \\
\hline 
MC axis orientation $(\theta_{ax}, \psi_{ax})$  & ($2^\circ$, $83^\circ$) & ($5\pm6^\circ$, $64\pm2^\circ$) & ($4^\circ$, $74^\circ$) \\ 

Impact parameter $(\, p_0 \,)^{\; b}$ & $-0.45$ & $-0.65$ & $-0.50$ \\

Eigenvalue ratio $(\, \lambda_2/\lambda_3 \, )^{\; c}$ & --- & $2.3\pm0.2$ & --- \\
Root-mean-square error $(\, E_{\rm rms} \, )^{\; d}$ & 0.32 & --- & --- \\ 

\hline
\hline

Azimuthal flux $(\, \phi^p_{\rm dir, out} / L_{\rm mc} \,)$  & $1.35\times10^{21}$ Mx/au & 1.5$\, \pm\, 0.2\times10^{21}$ Mx/au & $1.4\times10^{21}$ Mx/au \\

\hline
Percentage of flux erosion ($\phi_{\rm er}$) & $8\,\%$ & $29\pm13\,\%$ & $18\pm11\,\%$ \\

\hline
\multicolumn{2}{l}{Time of accumulated azimuthal flux imbalance$^{\, e}$} &  \multicolumn{2}{l}{$t^{*}_{\rm out}$ = 07:30 UT 26 June } \\

\hline
\multicolumn{2}{l}{Estimated \emph{start time} of the flux erosion$^{\, e}$} & \multicolumn{2}{l}{$t_{\rm rec}\sim$17:00 UT 24 June} \\

\hline
\multicolumn{2}{l}{Estimated \emph{start distance} of the flux erosion$^{\, e}$} & \multicolumn{2}{l}{ $r_{\rm rec}\sim$0.35 au } \\

\hline
\hline
 \multicolumn{4}{l}{$^{a}$ \small Nested-bootstrap MVA \citep[see][]{ruffenach2015statistical}.} \\
 \multicolumn{4}{l}{$^{b}$ \small A negative value means the spacecraft crosses south of the MC axis. } \\
 \multicolumn{4}{l}{$^{c}$ \small The intermediate ($\lambda_2$) to minimum ($\lambda_3$) eignenvalue ratio determined from MVA.} \\
 \multicolumn{4}{l}{$^{d}$ \small Defined as  $E_{\rm rms} =  \big( \sum_{i=1}^{N} \left[ \mathbf{B}^{\rm obs}(t_i) - \mathbf{B}^{\rm LFF}(t_i) \right]^{2} \big)^{1/2} / \big( N \max{ | \mathbf{B}^{\rm obs} |} \big)$ in \citet{marubashi2007long}.} \\
 \multicolumn{4}{l}{$^{e}$ \small Calculated using the mean( {LFF}, {nested-bootstrap MVA} ) values.} \\
\end{tabular}
\label{table:s2}
\end{table}
 
\section{Discussion and Conclusions}
\label{sec:disc}

The primary focus of this work was to investigate the initiation and formation of a stealth CME and its interaction with surrounding interplanetary structures. The spatiotemporal proximity between the sequence of three eruptions observed over STEREO-A's west limb, combined with the large-scale geometry of the PFSS coronal field extrapolation, strongly suggest a direct and causal magnetic coupling between the eruptions \citep{schrijver2013}.
In the \citet{torok2011} MHD simulation, their coupled, sympathetic eruptions from the adjacent flux systems of a coronal pseudostreamer were triggered by a prior eruption \textit{external} to the pseudostreamer arcades. In our study, CME \#0 plays an analogous role, removing a portion of the overlying (restraining) helmet streamer flux and disrupting the quasi-equilibrium force-balance of the energized, multipolar flux system. CMEs \#1 and \#2 then erupt sympathetically, in succession. CME \#1 starts in the northern side-lobe arcade, is seen to deflect south, and given the orientation of the PFSS arcades in Figure~\ref{f1}, likely propagates toward STEREO-A. CME \#2 starts in the southern side-lobe and is deflected northward \citep{palmerio2021predicting}. As discussed in \citet{lynch2013sympathetic}, each of the sympathetic CMEs' non-radial deflections are toward their overlying breakout current sheet which corresponds to the local magnetic pressure minimum (path of least resistance). More generally, there are many examples of mid- to high-latitude eruptions deflecting towards the HCS as they become SBO-CMEs \citep[e.g.,][]{kilpua2009deflect,panasenco2013,Zheng2020,Getachew2022}. Further numerical modeling is needed to confirm this multipolar sympathetic eruption scenario and to characterize the reconnection flux transfer between arcades during this multi-stage, SBO-CME event. 

At ${\sim}0.5$~AU, PSP witnessed the draping of IMF that reconnected with the SBO-CME (CME \#2) and eroded almost 20\% of its azimuthal flux. Analysing the MC's back region populated with reconnected field lines, we estimate that the reconnection might have initiated after 17:00~UT on 24 June at a heliospheric distance of ${\sim}0.35$~AU. A lower inclination (${\sim}29^\circ$) of plasma sheet behind the CME than the inclination (${\sim}41^\circ$) of plasma sheet in front of it indicates the draping of IMF about the MC having asymmetric, expanding, and non-circular FR structure. The asymmetry in magnetic field intensity has been quantified by $C_{B,t}$ \citep{janvier2019generic} and $C_{B,x}$ \citep{demoulin2020contribution} in temporal and spatial coordinates, respectively, where a negative value of $C_{B,t}$ and $C_{B,x}$ indicates a non-circular, expanding cross-section of a FR. In our study we find $C_{B,t} \sim C_{B,x} \sim -0.04$.

With the aid of coronal field extrapolations as well as remote and in-situ observations, we explain in this study the formation of a classic slow SBO-CME and examine the interplay with its surroundings in the inner heliosphere. By analyzing the PFSS geometry and the dynamics in low-coronal EUV and white-light imagery, we show that the CME followed two prior eruptions from nearby source regions and suggest coronal reconnection played a significant role in the eruption process. The CME's inner-heliospheric propagation resulted in a draping of the heliospheric field adjacent to the HCS about the CME flux rope. It forced the HCS to be inclined to the spacecraft orbital plane with a moderate and low inclination angle ahead and behind the CME, respectively, at a ${\sim}0.5$~AU heliospheric distance. Thus, this study provides important implications for the origin and interactions of a slow eruptive flux rope in the interplanetary medium and highlights the necessity of continuous off-limb observations away from the Sun--Earth line leading to a better exploration of the dynamics of stealthy CMEs in the low corona and inner heliosphere. 

\section*{Data Availability Statement}
Remote-sensing data from SDO, SOHO, and STEREO are openly available at the Virtual Solar Observatory (VSO; \url{https://sdac.virtualsolar.org/}) and STEREO Science Center website (SSC; \url{https://stereo-ssc.nascom.nasa.gov/}).
Additionally, this work made use of ESA's \textit{JHelioviewer} \cite{muller2017jhelioviewer} software.
PSP in-situ data are available at at NASA's Coordinated Data Analysis Web (CDAWeb; \url{https://cdaweb.gsfc.nasa.gov/index.html/}) database.
\section*{Author Contributions}
SP, BJL, SWG, EP and EK contributed to conception and design of the study. SP, BJL, SWG and EP performed the analysis. SP and BJL wrote the first draft of the manuscript. MLS organised the solar wind data. All authors contributed to manuscript revision, and approved the submitted version.

\section*{Acknowledgments}
S.~Pal and E.~Kilpua acknowledge the European Research Council (ERC) under the European Union's Horizon 2020 Research and Innovation Program Project SolMAG 724391. B.~Lynch acknowledges support from NSF AGS-1851945, NASA 80NSSC19K0088, NASA 80NSSC21K0731, and the HERMES Heliophysics DRIVE Center. S.~Good and E.~Kilpua acknowledge Academy of Finland Project 310445 (SMASH). E.~Palmerio acknowledges support from NASA's PSP-GI (no. 80NSSC22K0349) and HTMS (no. 80NSSC20K1274) programs. E.~Asvestari acknowledges Academy of Finland (Postdoctoral Grant no. 322455). S.~Pal thanks Dr.~Katsuhide~Marubashi for providing us with the linear force-free cylindrical model. We acknowledge the PSP/FIELDS, PSP/SWEAP, SDO/AIA, SOHO/LASCO, and STEREO/SECCHI teams.

\section*{Conflict of Interest Statement}
Author E. Palmerio is employed by Predictive Science Inc. The remaining authors declare that the research was conducted in the absence of any commercial or financial relationships that could be construed as a potential conflict of interest.

\bibliographystyle{Frontiers-Harvard} 
\bibliography{references.bib}

\begin{thebibliography}{83}
\providecommand{\natexlab}[1]{#1}
\expandafter\ifx\csname urlstyle\endcsname\relax
  \providecommand{\doi}[1]{doi:\discretionary{}{}{}#1}\else
  \providecommand{\doi}{doi:\discretionary{}{}{}\begingroup
  \urlstyle{rm}\Url}\fi
\providecommand{\selectlanguage}[1]{\relax}
\providecommand{\bibAnnoteFile}[1]{%
  \IfFileExists{#1}{\begin{quotation}\noindent\textsc{Key:} #1\\
  \textsc{Annotation:}\ \input{#1}\end{quotation}}{}}
\providecommand{\bibAnnote}[2]{%
  \begin{quotation}\noindent\textsc{Key:} #1\\
  \textsc{Annotation:}\ #2\end{quotation}}

\bibitem[{{Antiochos} et~al.(1999){Antiochos}, {DeVore}, and
  {Klimchuk}}]{antiochos1999}
{Antiochos}, S.~K., {DeVore}, C.~R., and {Klimchuk}, J.~A. (1999).
\newblock {A Model for Solar Coronal Mass Ejections}.
\newblock \emph{The Astrophysical Journal} 510, 485--493.
\newblock \doi{10.1086/306563}
\bibAnnoteFile{antiochos1999}

\bibitem[{{Bale} et~al.(2016){Bale}, {Goetz}, {Harvey}, {Turin}, {Bonnell},
  {Dudok de Wit} et~al.}]{bale2016fields}
{Bale}, S.~D., {Goetz}, K., {Harvey}, P.~R., {Turin}, P., {Bonnell}, J.~W.,
  {Dudok de Wit}, T., et~al. (2016).
\newblock {The FIELDS Instrument Suite for Solar Probe Plus. Measuring the
  Coronal Plasma and Magnetic Field, Plasma Waves and Turbulence, and Radio
  Signatures of Solar Transients}.
\newblock \emph{Space Science Reviews} 204, 49--82.
\newblock \doi{10.1007/s11214-016-0244-5}
\bibAnnoteFile{bale2016fields}

\bibitem[{{Brueckner} et~al.(1995){Brueckner}, {Howard}, {Koomen}, {Korendyke},
  {Michels}, {Moses} et~al.}]{brueckner1995large}
{Brueckner}, G.~E., {Howard}, R.~A., {Koomen}, M.~J., {Korendyke}, C.~M.,
  {Michels}, D.~J., {Moses}, J.~D., et~al. (1995).
\newblock {The Large Angle Spectroscopic Coronagraph (LASCO)}.
\newblock \emph{Solar Physics} 162, 357--402.
\newblock \doi{10.1007/BF00733434}
\bibAnnoteFile{brueckner1995large}

\bibitem[{{Burlaga}(1988)}]{burlaga1988magnetic}
{Burlaga}, L.~F. (1988).
\newblock {Magnetic clouds and force-free fields with constant alpha}.
\newblock \emph{Journal of Geophysical Research} 93, 7217--7224.
\newblock \doi{10.1029/JA093iA07p07217}
\bibAnnoteFile{burlaga1988magnetic}

\bibitem[{{Carcaboso} et~al.(2020){Carcaboso}, {G{\'o}mez-Herrero}, {Espinosa
  Lara}, {Hidalgo}, {Cernuda}, and
  {Rodr{\'\i}guez-Pacheco}}]{carcaboso2020characterisation}
{Carcaboso}, F., {G{\'o}mez-Herrero}, R., {Espinosa Lara}, F., {Hidalgo},
  M.~A., {Cernuda}, I., and {Rodr{\'\i}guez-Pacheco}, J. (2020).
\newblock {Characterisation of suprathermal electron pitch-angle distributions.
  Bidirectional and isotropic periods in solar wind}.
\newblock \emph{Astronomy \& Astrophysics} 635, A79.
\newblock \doi{10.1051/0004-6361/201936601}
\bibAnnoteFile{carcaboso2020characterisation}

\bibitem[{{Case} et~al.(2020){Case}, {Kasper}, {Stevens}, {Korreck}, {Paulson},
  {Daigneau} et~al.}]{case2020solar}
{Case}, A.~W., {Kasper}, J.~C., {Stevens}, M.~L., {Korreck}, K.~E., {Paulson},
  K., {Daigneau}, P., et~al. (2020).
\newblock {The Solar Probe Cup on the Parker Solar Probe}.
\newblock \emph{The Astrophysical Journal Supplement Series} 246, 43.
\newblock \doi{10.3847/1538-4365/ab5a7b}
\bibAnnoteFile{case2020solar}

\bibitem[{{Chollet} et~al.(2010){Chollet}, {Skoug}, {Steinberg}, {Crooker}, and
  {Giacalone}}]{chollet2010reconnection}
{Chollet}, E., {Skoug}, R., {Steinberg}, J., {Crooker}, N., and {Giacalone}, J.
  (2010).
\newblock {Reconnection and Disconnection: Observations of Suprathermal
  Electron Heat Flux Dropouts}.
\newblock In \emph{Twelfth International Solar Wind Conference}, eds.
  M.~{Maksimovic}, K.~{Issautier}, N.~{Meyer-Vernet}, M.~{Moncuquet}, and
  F.~{Pantellini}. vol. 1216 of \emph{American Institute of Physics Conference
  Series}, 600--603.
\newblock \doi{10.1063/1.3395937}
\bibAnnoteFile{chollet2010reconnection}

\bibitem[{{Crooker} et~al.(1996){Crooker}, {Burton}, {Siscoe}, {Kahler},
  {Gosling}, and {Smith}}]{crooker1996solar}
{Crooker}, N.~U., {Burton}, M.~E., {Siscoe}, G.~L., {Kahler}, S.~W., {Gosling},
  J.~T., and {Smith}, E.~J. (1996).
\newblock {Solar wind streamer belt structure}.
\newblock \emph{Journal of Geophysical Research} 101, 24331--24342.
\newblock \doi{10.1029/96JA02412}
\bibAnnoteFile{crooker1996solar}

\bibitem[{{Crooker} et~al.(2004){Crooker}, {Huang}, {Lamassa}, {Larson},
  {Kahler}, and {Spence}}]{crooker2004heliospheric}
{Crooker}, N.~U., {Huang}, C.~L., {Lamassa}, S.~M., {Larson}, D.~E., {Kahler},
  S.~W., and {Spence}, H.~E. (2004).
\newblock {Heliospheric plasma sheets}.
\newblock \emph{Journal of Geophysical Research} 109, A03107.
\newblock \doi{10.1029/2003JA010170}
\bibAnnoteFile{crooker2004heliospheric}

\bibitem[{{Crooker} et~al.(2003){Crooker}, {Larson}, {Kahler}, {Lamassa}, and
  {Spence}}]{crooker2003suprathermal}
{Crooker}, N.~U., {Larson}, D.~E., {Kahler}, S.~W., {Lamassa}, S.~M., and
  {Spence}, H.~E. (2003).
\newblock {Suprathermal electron isotropy in high-beta solar wind and its role
  in heat flux dropouts}.
\newblock \emph{Geophysical Research Letters} 30, 1619.
\newblock \doi{10.1029/2003GL017036}
\bibAnnoteFile{crooker2003suprathermal}

\bibitem[{{Dahlin} et~al.(2019){Dahlin}, {Antiochos}, and
  {DeVore}}]{dahlin2019model}
{Dahlin}, J.~T., {Antiochos}, S.~K., and {DeVore}, C.~R. (2019).
\newblock {A Model for Energy Buildup and Eruption Onset in Coronal Mass
  Ejections}.
\newblock \emph{The Astrophysical Journal} 879, 96.
\newblock \doi{10.3847/1538-4357/ab262a}
\bibAnnoteFile{dahlin2019model}

\bibitem[{{Dasso} et~al.(2005{\natexlab{a}}){Dasso}, {Gulisano}, {Mandrini},
  and {D{\'e}moulin}}]{dasso2005model}
{Dasso}, S., {Gulisano}, A.~M., {Mandrini}, C.~H., and {D{\'e}moulin}, P.
  (2005{\natexlab{a}}).
\newblock {Model-independent large-scale magnetohydrodynamic quantities in
  magnetic clouds}.
\newblock \emph{Advances in Space Research} 35, 2172--2177.
\newblock \doi{10.1016/j.asr.2005.03.054}
\bibAnnoteFile{dasso2005model}

\bibitem[{{Dasso} et~al.(2006){Dasso}, {Mandrini}, {D{\'e}moulin}, and
  {Luoni}}]{dasso2006model}
{Dasso}, S., {Mandrini}, C.~H., {D{\'e}moulin}, P., and {Luoni}, M.~L. (2006).
\newblock {A new model-independent method to compute magnetic helicity in
  magnetic clouds}.
\newblock \emph{Astronomy \& Astrophysics} 455, 349--359.
\newblock \doi{10.1051/0004-6361:20064806}
\bibAnnoteFile{dasso2006model}

\bibitem[{{Dasso} et~al.(2005{\natexlab{b}}){Dasso}, {Mandrini}, {Luoni},
  {Gulisano}, {Nakwacki}, {Pohjolainen} et~al.}]{dasso2005linking}
{Dasso}, S., {Mandrini}, C.~H., {Luoni}, M.~L., {Gulisano}, A.~M., {Nakwacki},
  M.~S., {Pohjolainen}, S., et~al. (2005{\natexlab{b}}).
\newblock {Linking Coronal to Heliospheric Magnetic Helicity: A New
  Model-Independent Technique to Compute Helicity in Magnetic Clouds}.
\newblock In \emph{Solar Wind 11/SOHO 16, Connecting Sun and Heliosphere}, eds.
  B.~{Fleck}, T.~H. {Zurbuchen}, and H.~{Lacoste}. vol. 592 of \emph{ESA
  Special Publication}, 605
\bibAnnoteFile{dasso2005linking}

\bibitem[{{Dasso} et~al.(2007){Dasso}, {Nakwacki}, {D{\'e}moulin}, and
  {Mandrini}}]{dasso2007progressive}
{Dasso}, S., {Nakwacki}, M.~S., {D{\'e}moulin}, P., and {Mandrini}, C.~H.
  (2007).
\newblock {Progressive Transformation of a Flux Rope to an ICME. Comparative
  Analysis Using the Direct and Fitted Expansion Methods}.
\newblock \emph{Solar Physics} 244, 115--137.
\newblock \doi{10.1007/s11207-007-9034-2}
\bibAnnoteFile{dasso2007progressive}

\bibitem[{D{\'e}moulin and Dasso(2009)}]{demoulin2009magnetic}
D{\'e}moulin, P. and Dasso, S. (2009).
\newblock Magnetic cloud models with bent and oblate cross-section boundaries.
\newblock \emph{Astronomy \& Astrophysics} 507, 969--980
\bibAnnoteFile{demoulin2009magnetic}

\bibitem[{{D{\'e}moulin} et~al.(2020){D{\'e}moulin}, {Dasso}, {Lanabere}, and
  {Janvier}}]{demoulin2020contribution}
{D{\'e}moulin}, P., {Dasso}, S., {Lanabere}, V., and {Janvier}, M. (2020).
\newblock {Contribution of the ageing effect to the observed asymmetry of
  interplanetary magnetic clouds}.
\newblock \emph{Astronomy \& Astrophysics} 639, A6.
\newblock \doi{10.1051/0004-6361/202038077}
\bibAnnoteFile{demoulin2020contribution}

\bibitem[{{DeVore} and {Antiochos}(2008)}]{devore2008homologous}
{DeVore}, C.~R. and {Antiochos}, S.~K. (2008).
\newblock {Homologous Confined Filament Eruptions via Magnetic Breakout}.
\newblock \emph{The Astrophysical Journal} 680, 740--756.
\newblock \doi{10.1086/588011}
\bibAnnoteFile{devore2008homologous}

\bibitem[{{Domingo} et~al.(1995){Domingo}, {Fleck}, and
  {Poland}}]{domingo1995soho}
{Domingo}, V., {Fleck}, B., and {Poland}, A.~I. (1995).
\newblock {The SOHO Mission: an Overview}.
\newblock \emph{Solar Physics} 162, 1--37.
\newblock \doi{10.1007/BF00733425}
\bibAnnoteFile{domingo1995soho}

\bibitem[{{Fox} et~al.(2016){Fox}, {Velli}, {Bale}, {Decker}, {Driesman},
  {Howard} et~al.}]{fox2016solar}
{Fox}, N.~J., {Velli}, M.~C., {Bale}, S.~D., {Decker}, R., {Driesman}, A.,
  {Howard}, R.~A., et~al. (2016).
\newblock {The Solar Probe Plus Mission: Humanity's First Visit to Our Star}.
\newblock \emph{Space Science Reviews} 204, 7--48.
\newblock \doi{10.1007/s11214-015-0211-6}
\bibAnnoteFile{fox2016solar}

\bibitem[{{Getachew} et~al.(2022){Getachew}, {McComas}, {Joyce}, {Palmerio},
  {Christian}, {Cohen} et~al.}]{Getachew2022}
{Getachew}, T., {McComas}, D.~J., {Joyce}, C.~J., {Palmerio}, E., {Christian},
  E.~R., {Cohen}, C.~M.~S., et~al. (2022).
\newblock {PSP/IS{\ensuremath{\odot}}IS Observation of a Solar Energetic
  Particle Event Associated with a Streamer Blowout Coronal Mass Ejection
  during Encounter 6}.
\newblock \emph{The Astrophysical Journal} 925, 212.
\newblock \doi{10.3847/1538-4357/ac408f}
\bibAnnoteFile{Getachew2022}

\bibitem[{{Gosling} et~al.(1974){Gosling}, {Hildner}, {MacQueen}, {Munro},
  {Poland}, and {Ross}}]{gosling1974mass}
{Gosling}, J.~T., {Hildner}, E., {MacQueen}, R.~M., {Munro}, R.~H., {Poland},
  A.~I., and {Ross}, C.~L. (1974).
\newblock {Mass ejections from the Sun: A view from Skylab}.
\newblock \emph{Journal of Geophysical Research} 79, 4581.
\newblock \doi{10.1029/JA079i031p04581}
\bibAnnoteFile{gosling1974mass}

\bibitem[{{Gosling} and {McComas}(1987)}]{gosling1987field}
{Gosling}, J.~T. and {McComas}, D.~J. (1987).
\newblock {Field line draping about fast coronal mass ejecta: A source of
  strong out-of-the-ecliptic interplanetary magnetic fields}.
\newblock \emph{Geophysical Research Letters} 14, 355--358.
\newblock \doi{10.1029/GL014i004p00355}
\bibAnnoteFile{gosling1987field}

\bibitem[{{Gosling} et~al.(2005{\natexlab{a}}){Gosling}, {Skoug}, {McComas},
  and {Smith}}]{gosling2005direct}
{Gosling}, J.~T., {Skoug}, R.~M., {McComas}, D.~J., and {Smith}, C.~W.
  (2005{\natexlab{a}}).
\newblock {Direct evidence for magnetic reconnection in the solar wind near 1
  AU}.
\newblock \emph{Journal of Geophysical Research} 110, A01107.
\newblock \doi{10.1029/2004JA010809}
\bibAnnoteFile{gosling2005direct}

\bibitem[{{Gosling} et~al.(2005{\natexlab{b}}){Gosling}, {Skoug}, {McComas},
  and {Smith}}]{gosling2005magnetic}
{Gosling}, J.~T., {Skoug}, R.~M., {McComas}, D.~J., and {Smith}, C.~W.
  (2005{\natexlab{b}}).
\newblock {Magnetic disconnection from the Sun: Observations of a reconnection
  exhaust in the solar wind at the heliospheric current sheet}.
\newblock \emph{Geophysical Research Letters} 32, L05105.
\newblock \doi{10.1029/2005GL022406}
\bibAnnoteFile{gosling2005magnetic}

\bibitem[{{Hidalgo} et~al.(2000){Hidalgo}, {Cid}, {Medina}, and
  {Vi{\~n}as}}]{hidalgo2000model}
{Hidalgo}, M.~A., {Cid}, C., {Medina}, J., and {Vi{\~n}as}, A.~F. (2000).
\newblock {A new model for the topology of magnetic clouds in the solar wind}.
\newblock \emph{Solar Physics} 194, 165--174.
\newblock \doi{10.1023/A:1005206107017}
\bibAnnoteFile{hidalgo2000model}

\bibitem[{{Howard} et~al.(2008){Howard}, {Moses}, {Vourlidas}, {Newmark},
  {Socker}, {Plunkett} et~al.}]{howard2008sun}
{Howard}, R.~A., {Moses}, J.~D., {Vourlidas}, A., {Newmark}, J.~S., {Socker},
  D.~G., {Plunkett}, S.~P., et~al. (2008).
\newblock {Sun Earth Connection Coronal and Heliospheric Investigation
  (SECCHI)}.
\newblock \emph{Space Science Reviews} 136, 67--115.
\newblock \doi{10.1007/s11214-008-9341-4}
\bibAnnoteFile{howard2008sun}

\bibitem[{{Howard} and {Harrison}(2013)}]{howard2013stealth}
{Howard}, T.~A. and {Harrison}, R.~A. (2013).
\newblock {Stealth Coronal Mass Ejections: A Perspective}.
\newblock \emph{Solar Physics} 285, 269--280.
\newblock \doi{10.1007/s11207-012-0217-0}
\bibAnnoteFile{howard2013stealth}

\bibitem[{{Illing} and {Hundhausen}(1983)}]{illing1983possible}
{Illing}, R.~M.~E. and {Hundhausen}, A.~J. (1983).
\newblock {Possible observation of a disconnected magnetic structure in a
  coronal transient}.
\newblock \emph{Journal of Geophysical Research} 88, 10210--10214.
\newblock \doi{10.1029/JA088iA12p10210}
\bibAnnoteFile{illing1983possible}

\bibitem[{{Janvier} et~al.(2019){Janvier}, {Winslow}, {Good}, {Bonhomme},
  {D{\'e}moulin}, {Dasso} et~al.}]{janvier2019generic}
{Janvier}, M., {Winslow}, R.~M., {Good}, S., {Bonhomme}, E., {D{\'e}moulin},
  P., {Dasso}, S., et~al. (2019).
\newblock {Generic Magnetic Field Intensity Profiles of Interplanetary Coronal
  Mass Ejections at Mercury, Venus, and Earth From Superposed Epoch Analyses}.
\newblock \emph{Journal of Geophysical Research: Space Physics} 124, 812--836.
\newblock \doi{10.1029/2018JA025949}
\bibAnnoteFile{janvier2019generic}

\bibitem[{{Kahler} et~al.(1996){Kahler}, {Crooker}, and
  {Gosling}}]{kahler1996topology}
{Kahler}, S.~W., {Crooker}, N.~U., and {Gosling}, J.~T. (1996).
\newblock {The topology of intrasector reversals of the interplanetary magnetic
  field}.
\newblock \emph{Journal of Geophysical Research} 101, 24373--24382.
\newblock \doi{10.1029/96JA02232}
\bibAnnoteFile{kahler1996topology}

\bibitem[{{Kahler} and {Lin}(1994)}]{kahler1994determination}
{Kahler}, S.~W. and {Lin}, R.~P. (1994).
\newblock {The determination of interplanetary magnetic field polarities around
  sector boundaries using E $>$ 2 keV electrons}.
\newblock \emph{Geophysical Research Letters} 21, 1575--1578.
\newblock \doi{10.1029/94GL01362}
\bibAnnoteFile{kahler1994determination}

\bibitem[{{Kaiser} et~al.(2008){Kaiser}, {Kucera}, {Davila}, {St. Cyr},
  {Guhathakurta}, and {Christian}}]{kaiser2008stereo}
{Kaiser}, M.~L., {Kucera}, T.~A., {Davila}, J.~M., {St. Cyr}, O.~C.,
  {Guhathakurta}, M., and {Christian}, E. (2008).
\newblock {The STEREO Mission: An Introduction}.
\newblock \emph{Space Science Reviews} 136, 5--16.
\newblock \doi{10.1007/s11214-007-9277-0}
\bibAnnoteFile{kaiser2008stereo}

\bibitem[{{Kasper} et~al.(2016){Kasper}, {Abiad}, {Austin}, {Balat-Pichelin},
  {Bale}, {Belcher} et~al.}]{kasper2016solar}
{Kasper}, J.~C., {Abiad}, R., {Austin}, G., {Balat-Pichelin}, M., {Bale},
  S.~D., {Belcher}, J.~W., et~al. (2016).
\newblock {Solar Wind Electrons Alphas and Protons (SWEAP) Investigation:
  Design of the Solar Wind and Coronal Plasma Instrument Suite for Solar Probe
  Plus}.
\newblock \emph{Space Science Reviews} 204, 131--186.
\newblock \doi{10.1007/s11214-015-0206-3}
\bibAnnoteFile{kasper2016solar}

\bibitem[{{Kawano} and {Higuchi}(1995)}]{kawano1995bootstrap}
{Kawano}, H. and {Higuchi}, T. (1995).
\newblock {The bootstrap method in space physics: Error estimation for the
  minimum variance analysis}.
\newblock \emph{Geophysical Research Letters} 22, 307--310.
\newblock \doi{10.1029/94GL02969}
\bibAnnoteFile{kawano1995bootstrap}

\bibitem[{{Kay} et~al.(2015){Kay}, {Opher}, and {Evans}}]{kay2015global}
{Kay}, C., {Opher}, M., and {Evans}, R.~M. (2015).
\newblock {Global Trends of CME Deflections Based on CME and Solar Parameters}.
\newblock \emph{The Astrophysical Journal} 805, 168.
\newblock \doi{10.1088/0004-637X/805/2/168}
\bibAnnoteFile{kay2015global}

\bibitem[{{Kilpua} et~al.(2013){Kilpua}, {Isavnin}, {Vourlidas}, {Koskinen},
  and {Rodriguez}}]{kilpua2013relationship}
{Kilpua}, E.~K.~J., {Isavnin}, A., {Vourlidas}, A., {Koskinen}, H.~E.~J., and
  {Rodriguez}, L. (2013).
\newblock {On the relationship between interplanetary coronal mass ejections
  and magnetic clouds}.
\newblock \emph{Annales Geophysicae} 31, 1251--1265.
\newblock \doi{10.5194/angeo-31-1251-2013}
\bibAnnoteFile{kilpua2013relationship}

\bibitem[{{Kilpua} et~al.(2014){Kilpua}, {Mierla}, {Zhukov}, {Rodriguez},
  {Vourlidas}, and {Wood}}]{kilpua2014solar}
{Kilpua}, E.~K.~J., {Mierla}, M., {Zhukov}, A.~N., {Rodriguez}, L.,
  {Vourlidas}, A., and {Wood}, B. (2014).
\newblock {Solar Sources of Interplanetary Coronal Mass Ejections During the
  Solar Cycle 23/24 Minimum}.
\newblock \emph{Solar Physics} 289, 3773--3797.
\newblock \doi{10.1007/s11207-014-0552-4}
\bibAnnoteFile{kilpua2014solar}

\bibitem[{{Kilpua} et~al.(2009){Kilpua}, {Pomoell}, {Vourlidas}, {Vainio},
  {Luhmann}, {Li} et~al.}]{kilpua2009deflect}
{Kilpua}, E.~K.~J., {Pomoell}, J., {Vourlidas}, A., {Vainio}, R., {Luhmann},
  J., {Li}, Y., et~al. (2009).
\newblock {STEREO observations of interplanetary coronal mass ejections and
  prominence deflection during solar minimum period}.
\newblock \emph{Annales Geophysicae} 27, 4491--4503.
\newblock \doi{10.5194/angeo-27-4491-2009}
\bibAnnoteFile{kilpua2009deflect}

\bibitem[{{Korreck} et~al.(2020){Korreck}, {Szabo}, {Nieves Chinchilla},
  {Lavraud}, {Luhmann}, {Niembro} et~al.}]{korreck2020source}
{Korreck}, K.~E., {Szabo}, A., {Nieves Chinchilla}, T., {Lavraud}, B.,
  {Luhmann}, J., {Niembro}, T., et~al. (2020).
\newblock {Source and Propagation of a Streamer Blowout Coronal Mass Ejection
  Observed by the Parker Solar Probe}.
\newblock \emph{The Astrophysical Journal Supplement Series} 246, 69.
\newblock \doi{10.3847/1538-4365/ab6ff9}
\bibAnnoteFile{korreck2020source}

\bibitem[{{Lavraud} et~al.(2020){Lavraud}, {Fargette}, {R{\'e}ville}, {Szabo},
  {Huang}, {Rouillard} et~al.}]{lavraud2020heliospheric}
{Lavraud}, B., {Fargette}, N., {R{\'e}ville}, V., {Szabo}, A., {Huang}, J.,
  {Rouillard}, A.~P., et~al. (2020).
\newblock {The Heliospheric Current Sheet and Plasma Sheet during Parker Solar
  Probe's First Orbit}.
\newblock \emph{The Astrophysical Journal Letters} 894, L19.
\newblock \doi{10.3847/2041-8213/ab8d2d}
\bibAnnoteFile{lavraud2020heliospheric}

\bibitem[{{Lemen} et~al.(2012){Lemen}, {Title}, {Akin}, {Boerner}, {Chou},
  {Drake} et~al.}]{lemen2012atmospheric}
{Lemen}, J.~R., {Title}, A.~M., {Akin}, D.~J., {Boerner}, P.~F., {Chou}, C.,
  {Drake}, J.~F., et~al. (2012).
\newblock {The Atmospheric Imaging Assembly (AIA) on the Solar Dynamics
  Observatory (SDO)}.
\newblock \emph{Solar Physics} 275, 17--40.
\newblock \doi{10.1007/s11207-011-9776-8}
\bibAnnoteFile{lemen2012atmospheric}

\bibitem[{{Lepping} et~al.(1990){Lepping}, {Jones}, and
  {Burlaga}}]{lepping1990magnetic}
{Lepping}, R.~P., {Jones}, J.~A., and {Burlaga}, L.~F. (1990).
\newblock {Magnetic field structure of interplanetary magnetic clouds at 1 AU}.
\newblock \emph{Journal of Geophysical Research} 95, 11957--11965.
\newblock \doi{10.1029/JA095iA08p11957}
\bibAnnoteFile{lepping1990magnetic}

\bibitem[{{Lepping} et~al.(1996{\natexlab{a}}){Lepping}, {Szabo}, {Peredo}, and
  {Campbell}}]{lepping1996summary}
{Lepping}, R.~P., {Szabo}, A., {Peredo}, M., and {Campbell}, A.
  (1996{\natexlab{a}}).
\newblock {Summary of heliospheric current and plasma sheet studies: WIND
  observations}.
\newblock In \emph{Proceedings of the eigth International solar wind
  Conference: Solar wind eight}, eds. D.~{Winterhalter}, J.~T. {Gosling}, S.~R.
  {Habbal}, W.~S. {Kurth}, and M.~{Neugebauer}. vol. 382 of \emph{American
  Institute of Physics Conference Series}, 526--529.
\newblock \doi{10.1063/1.51505}
\bibAnnoteFile{lepping1996summary}

\bibitem[{{Lepping} et~al.(1996{\natexlab{b}}){Lepping}, {Szabo}, {Peredo}, and
  {Hoeksema}}]{lepping1996large}
{Lepping}, R.~P., {Szabo}, A., {Peredo}, M., and {Hoeksema}, J.~T.
  (1996{\natexlab{b}}).
\newblock {Large-scale properties and solar connection of the heliospheric
  current and plasma sheets: WIND observations}.
\newblock \emph{Geophysical Research Letters} 23, 1199--1202.
\newblock \doi{10.1029/96GL00658}
\bibAnnoteFile{lepping1996large}

\bibitem[{{Li} et~al.(2008){Li}, {Lynch}, {Stenborg}, {Luhmann}, {Huttunen},
  {Welsch} et~al.}]{li2008solar}
{Li}, Y., {Lynch}, B.~J., {Stenborg}, G., {Luhmann}, J.~G., {Huttunen},
  K.~E.~J., {Welsch}, B.~T., et~al. (2008).
\newblock {The Solar Magnetic Field and Coronal Dynamics of the Eruption on
  2007 May 19}.
\newblock \emph{The Astrophysical Journal Letters} 681, L37.
\newblock \doi{10.1086/590340}
\bibAnnoteFile{li2008solar}

\bibitem[{{Lynch} and {Edmondson}(2013)}]{lynch2013sympathetic}
{Lynch}, B.~J. and {Edmondson}, J.~K. (2013).
\newblock {Sympathetic Magnetic Breakout Coronal Mass Ejections from
  Pseudostreamers}.
\newblock \emph{The Astrophysical Journal} 764, 87.
\newblock \doi{10.1088/0004-637X/764/1/87}
\bibAnnoteFile{lynch2013sympathetic}

\bibitem[{{Lynch} et~al.(2016){Lynch}, {Masson}, {Li}, {DeVore}, {Luhmann},
  {Antiochos} et~al.}]{lynch2016model}
{Lynch}, B.~J., {Masson}, S., {Li}, Y., {DeVore}, C.~R., {Luhmann}, J.~G.,
  {Antiochos}, S.~K., et~al. (2016).
\newblock {A model for stealth coronal mass ejections}.
\newblock \emph{Journal of Geophysical Research: Space Physics} 121,
  10,677--10,697.
\newblock \doi{10.1002/2016JA023432}
\bibAnnoteFile{lynch2016model}

\bibitem[{{Ma} et~al.(2010){Ma}, {Attrill}, {Golub}, and
  {Lin}}]{ma2010statistical}
{Ma}, S., {Attrill}, G.~D.~R., {Golub}, L., and {Lin}, J. (2010).
\newblock {Statistical Study of Coronal Mass Ejections With and Without
  Distinct Low Coronal Signatures}.
\newblock \emph{The Astrophysical Journal} 722, 289--301.
\newblock \doi{10.1088/0004-637X/722/1/289}
\bibAnnoteFile{ma2010statistical}

\bibitem[{{Marubashi} and {Lepping}(2007)}]{marubashi2007long}
{Marubashi}, K. and {Lepping}, R.~P. (2007).
\newblock {Long-duration magnetic clouds: a comparison of analyses using torus-
  and cylinder-shaped flux rope models}.
\newblock \emph{Annales Geophysicae} 25, 2453--2477.
\newblock \doi{10.5194/angeo-25-2453-2007}
\bibAnnoteFile{marubashi2007long}

\bibitem[{{McComas} et~al.(1994){McComas}, {Gosling}, {Hammond}, {Moldwin},
  {Phillips}, and {Forsyth}}]{mccomas1994magnetic}
{McComas}, D.~J., {Gosling}, J.~T., {Hammond}, C.~M., {Moldwin}, M.~B.,
  {Phillips}, J.~L., and {Forsyth}, R.~J. (1994).
\newblock {Magnetic reconnection ahead of a coronal mass ejection}.
\newblock \emph{Geophysical Research Letters} 21, 1751--1754.
\newblock \doi{10.1029/94GL01077}
\bibAnnoteFile{mccomas1994magnetic}

\bibitem[{{McComas} et~al.(1989){McComas}, {Gosling}, {Phillips}, {Bame},
  {Luhmann}, and {Smith}}]{mccomas1989electron}
{McComas}, D.~J., {Gosling}, J.~T., {Phillips}, J.~L., {Bame}, S.~J.,
  {Luhmann}, J.~G., and {Smith}, E.~J. (1989).
\newblock {Electron heat flux dropouts in the solar wind: evidence for
  interplanetary magnetic field reconnection?}
\newblock \emph{Journal of Geophysical Research} 94, 6907--6916.
\newblock \doi{10.1029/JA094iA06p06907}
\bibAnnoteFile{mccomas1989electron}

\bibitem[{{McComas} et~al.(1988){McComas}, {Gosling}, {Winterhalter}, and
  {Smith}}]{mccomas1988interplanetary}
{McComas}, D.~J., {Gosling}, J.~T., {Winterhalter}, D., and {Smith}, E.~J.
  (1988).
\newblock {Interplanetary magnetic field draping about fast coronal mass ejecta
  in the outer heliosphere}.
\newblock \emph{Journal of Geophysical Research} 93, 2519--2526.
\newblock \doi{10.1029/JA093iA04p02519}
\bibAnnoteFile{mccomas1988interplanetary}

\bibitem[{{M{\"o}stl} et~al.(2022){M{\"o}stl}, {Weiss}, {Reiss}, {Amerstorfer},
  {Bailey}, {Hinterreiter} et~al.}]{mostl2021multipoint}
{M{\"o}stl}, C., {Weiss}, A.~J., {Reiss}, M.~A., {Amerstorfer}, T., {Bailey},
  R.~L., {Hinterreiter}, J., et~al. (2022).
\newblock {Multipoint Interplanetary Coronal Mass Ejections Observed with Solar
  Orbiter, BepiColombo, Parker Solar Probe, Wind, and STEREO-A}.
\newblock \emph{The Astrophysical Journal Letters} 924, L6.
\newblock \doi{10.3847/2041-8213/ac42d0}
\bibAnnoteFile{mostl2021multipoint}

\bibitem[{{M{\"u}ller} et~al.(2017){M{\"u}ller}, {Nicula}, {Felix},
  {Verstringe}, {Bourgoignie}, {Csillaghy} et~al.}]{muller2017jhelioviewer}
{M{\"u}ller}, D., {Nicula}, B., {Felix}, S., {Verstringe}, F., {Bourgoignie},
  B., {Csillaghy}, A., et~al. (2017).
\newblock {JHelioviewer. Time-dependent 3D visualisation of solar and
  heliospheric data}.
\newblock \emph{Astronomy \& Astrophysics} 606, A10.
\newblock \doi{10.1051/0004-6361/201730893}
\bibAnnoteFile{muller2017jhelioviewer}

\bibitem[{{Nitta} et~al.(2021){Nitta}, {Mulligan}, {Kilpua}, {Lynch}, {Mierla},
  {O'Kane} et~al.}]{nitta2021}
{Nitta}, N.~V., {Mulligan}, T., {Kilpua}, E. K.~J., {Lynch}, B.~J., {Mierla},
  M., {O'Kane}, J., et~al. (2021).
\newblock {Understanding the Origins of Problem Geomagnetic Storms Associated
  with ``Stealth'' Coronal Mass Ejections}.
\newblock \emph{Space Science Reviews} 217, 82.
\newblock \doi{10.1007/s11214-021-00857-0}
\bibAnnoteFile{nitta2021}

\bibitem[{{O'Kane} et~al.(2021){O'Kane}, {Green}, {Davies}, {M{\"o}stl},
  {Hinterreiter}, {Freiherr von Forstner} et~al.}]{okane2021solar}
{O'Kane}, J., {Green}, L.~M., {Davies}, E.~E., {M{\"o}stl}, C., {Hinterreiter},
  J., {Freiherr von Forstner}, J.~L., et~al. (2021).
\newblock {Solar origins of a strong stealth CME detected by Solar Orbiter}.
\newblock \emph{Astronomy \& Astrophysics} 656, L6.
\newblock \doi{10.1051/0004-6361/202140622}
\bibAnnoteFile{okane2021solar}

\bibitem[{Pal(2022)}]{PAL2021}
Pal, S. (2022).
\newblock Uncovering the process that transports magnetic helicity to coronal
  mass ejection flux ropes.
\newblock \emph{Advances in Space Research} in press.
\newblock \doi{10.1016/j.asr.2021.11.013}
\bibAnnoteFile{PAL2021}

\bibitem[{{Pal} et~al.(2020){Pal}, {Dash}, and {Nandy}}]{pal2020flux}
{Pal}, S., {Dash}, S., and {Nandy}, D. (2020).
\newblock {Flux Erosion of Magnetic Clouds by Reconnection With the Sun's Open
  Flux}.
\newblock \emph{Geophysical Research Letters} 47, e2019GL086372.
\newblock \doi{10.1029/2019GL086372}
\bibAnnoteFile{pal2020flux}

\bibitem[{{Pal} et~al.(2021){Pal}, {Kilpua}, {Good}, {Pomoell}, and
  {Price}}]{pal2021uncovering}
{Pal}, S., {Kilpua}, E., {Good}, S., {Pomoell}, J., and {Price}, D.~J. (2021).
\newblock {Uncovering erosion effects on magnetic flux rope twist}.
\newblock \emph{Astronomy \& Astrophysics} 650, A176.
\newblock \doi{10.1051/0004-6361/202040070}
\bibAnnoteFile{pal2021uncovering}

\bibitem[{{Palmerio} et~al.(2021{\natexlab{a}}){Palmerio}, {Kay}, {Al-Haddad},
  {Lynch}, {Yu}, {Stevens} et~al.}]{palmerio2021predicting}
{Palmerio}, E., {Kay}, C., {Al-Haddad}, N., {Lynch}, B.~J., {Yu}, W.,
  {Stevens}, M.~L., et~al. (2021{\natexlab{a}}).
\newblock {Predicting the Magnetic Fields of a Stealth CME Detected by Parker
  Solar Probe at 0.5 au}.
\newblock \emph{The Astrophysical Journal} 920, 65.
\newblock \doi{10.3847/1538-4357/ac25f4}
\bibAnnoteFile{palmerio2021predicting}

\bibitem[{{Palmerio} et~al.(2021{\natexlab{b}}){Palmerio}, {Nitta}, {Mulligan},
  {Mierla}, {O'Kane}, {Richardson} et~al.}]{palmerio2021investigating}
{Palmerio}, E., {Nitta}, N.~V., {Mulligan}, T., {Mierla}, M., {O'Kane}, J.,
  {Richardson}, I.~G., et~al. (2021{\natexlab{b}}).
\newblock {Investigating Remote-sensing Techniques to Reveal Stealth Coronal
  Mass Ejections}.
\newblock \emph{Frontiers in Astronomy and Space Sciences} 8, 695966.
\newblock \doi{10.3389/fspas.2021.695966}
\bibAnnoteFile{palmerio2021investigating}

\bibitem[{{Panasenco} et~al.(2013){Panasenco}, {Martin}, {Velli}, and
  {Vourlidas}}]{panasenco2013}
{Panasenco}, O., {Martin}, S.~F., {Velli}, M., and {Vourlidas}, A. (2013).
\newblock {Origins of Rolling, Twisting, and Non-radial Propagation of Eruptive
  Solar Events}.
\newblock \emph{Solar Physics} 287, 391--413.
\newblock \doi{10.1007/s11207-012-0194-3}
\bibAnnoteFile{panasenco2013}

\bibitem[{{Pesnell} et~al.(2012){Pesnell}, {Thompson}, and
  {Chamberlin}}]{pesnell2012solar}
{Pesnell}, W.~D., {Thompson}, B.~J., and {Chamberlin}, P.~C. (2012).
\newblock {The Solar Dynamics Observatory (SDO)}.
\newblock \emph{Solar Physics} 275, 3--15.
\newblock \doi{10.1007/s11207-011-9841-3}
\bibAnnoteFile{pesnell2012solar}

\bibitem[{Phan et~al.(2020)Phan, Bale, Eastwood, Lavraud, Drake, Oieroset
  et~al.}]{phan2020parker}
Phan, T., Bale, S., Eastwood, J., Lavraud, B., Drake, J., Oieroset, M., et~al.
  (2020).
\newblock Parker solar probe in situ observations of magnetic reconnection
  exhausts during encounter 1.
\newblock \emph{The Astrophysical Journal Supplement Series} 246, 34
\bibAnnoteFile{phan2020parker}

\bibitem[{{Robbrecht} et~al.(2009){Robbrecht}, {Patsourakos}, and
  {Vourlidas}}]{robbrecht2009trace}
{Robbrecht}, E., {Patsourakos}, S., and {Vourlidas}, A. (2009).
\newblock {No Trace Left Behind: STEREO Observation of a Coronal Mass Ejection
  Without Low Coronal Signatures}.
\newblock \emph{The Astrophysical Journal} 701, 283--291.
\newblock \doi{10.1088/0004-637X/701/1/283}
\bibAnnoteFile{robbrecht2009trace}

\bibitem[{{Ruffenach} et~al.(2015){Ruffenach}, {Lavraud}, {Farrugia},
  {D{\'e}moulin}, {Dasso}, {Owens} et~al.}]{ruffenach2015statistical}
{Ruffenach}, A., {Lavraud}, B., {Farrugia}, C.~J., {D{\'e}moulin}, P., {Dasso},
  S., {Owens}, M.~J., et~al. (2015).
\newblock {Statistical study of magnetic cloud erosion by magnetic
  reconnection}.
\newblock \emph{Journal of Geophysical Research: Space Physics} 120, 43--60.
\newblock \doi{10.1002/2014JA020628}
\bibAnnoteFile{ruffenach2015statistical}

\bibitem[{{Scherrer} et~al.(2012){Scherrer}, {Schou}, {Bush}, {Kosovichev},
  {Bogart}, {Hoeksema} et~al.}]{scherrer2012helioseismic}
{Scherrer}, P.~H., {Schou}, J., {Bush}, R.~I., {Kosovichev}, A.~G., {Bogart},
  R.~S., {Hoeksema}, J.~T., et~al. (2012).
\newblock {The Helioseismic and Magnetic Imager (HMI) Investigation for the
  Solar Dynamics Observatory (SDO)}.
\newblock \emph{Solar Physics} 275, 207--227.
\newblock \doi{10.1007/s11207-011-9834-2}
\bibAnnoteFile{scherrer2012helioseismic}

\bibitem[{{Schrijver} et~al.(2013){Schrijver}, {Title}, {Yeates}, and
  {DeRosa}}]{schrijver2013}
{Schrijver}, C.~J., {Title}, A.~M., {Yeates}, A.~R., and {DeRosa}, M.~L.
  (2013).
\newblock {Pathways of Large-scale Magnetic Couplings between Solar Coronal
  Events}.
\newblock \emph{The Astrophysical Journal} 773, 93.
\newblock \doi{10.1088/0004-637X/773/2/93}
\bibAnnoteFile{schrijver2013}

\bibitem[{{Sheeley} et~al.(1982){Sheeley}, {Howard}, {Koomen}, {Michels},
  {Harvey}, and {Harvey}}]{sheeley1982observations}
{Sheeley}, J., N.~R., {Howard}, R.~A., {Koomen}, M.~J., {Michels}, D.~J.,
  {Harvey}, K.~L., and {Harvey}, J.~W. (1982).
\newblock {Observations of coronal structure during sunspot maximum}.
\newblock \emph{Space Science Reviews} 33, 219--231.
\newblock \doi{10.1007/BF00213255}
\bibAnnoteFile{sheeley1982observations}

\bibitem[{{Sheeley} et~al.(1999){Sheeley}, {Walters}, {Wang}, and
  {Howard}}]{sheeley1999}
{Sheeley}, J., N.~R., {Walters}, J.~H., {Wang}, Y.~M., and {Howard}, R.~A.
  (1999).
\newblock {Continuous tracking of coronal outflows: Two kinds of coronal mass
  ejections}.
\newblock \emph{Journal of Geophysical Research} 104, 24739--24768.
\newblock \doi{10.1029/1999JA900308}
\bibAnnoteFile{sheeley1999}

\bibitem[{{Shen} et~al.(2012){Shen}, {Liu}, and {Su}}]{2012ApJ...750...12S}
{Shen}, Y., {Liu}, Y., and {Su}, J. (2012).
\newblock {Sympathetic Partial and Full Filament Eruptions Observed in One
  Solar Breakout Event}.
\newblock \emph{The Astrophysical Journal} 750, 12.
\newblock \doi{10.1088/0004-637X/750/1/12}
\bibAnnoteFile{2012ApJ...750...12S}

\bibitem[{{Smith}(2001)}]{smith2001heliospheric}
{Smith}, E.~J. (2001).
\newblock {The heliospheric current sheet}.
\newblock \emph{Journal of Geophysical Research} 106, 15819--15832.
\newblock \doi{10.1029/2000JA000120}
\bibAnnoteFile{smith2001heliospheric}

\bibitem[{{Sonnerup} and {Cahill}(1967)}]{sonnerup1967magnetopause}
{Sonnerup}, B.~U.~O. and {Cahill}, J., L.~J. (1967).
\newblock {Magnetopause Structure and Attitude from Explorer 12 Observations}.
\newblock \emph{Journal of Geophysical Research} 72, 171.
\newblock \doi{10.1029/JZ072i001p00171}
\bibAnnoteFile{sonnerup1967magnetopause}

\bibitem[{{Thernisien} et~al.(2006){Thernisien}, {Howard}, and
  {Vourlidas}}]{thernisien2006modeling}
{Thernisien}, A.~F.~R., {Howard}, R.~A., and {Vourlidas}, A. (2006).
\newblock {Modeling of Flux Rope Coronal Mass Ejections}.
\newblock \emph{The Astrophysical Journal} 652, 763--773.
\newblock \doi{10.1086/508254}
\bibAnnoteFile{thernisien2006modeling}

\bibitem[{{T{\"o}r{\"o}k} et~al.(2011){T{\"o}r{\"o}k}, {Panasenco}, {Titov},
  {Miki{\'c}}, {Reeves}, {Velli} et~al.}]{torok2011}
{T{\"o}r{\"o}k}, T., {Panasenco}, O., {Titov}, V.~S., {Miki{\'c}}, Z.,
  {Reeves}, K.~K., {Velli}, M., et~al. (2011).
\newblock {A Model for Magnetically Coupled Sympathetic Eruptions}.
\newblock \emph{The Astrophysical Journal Letters} 739, L63.
\newblock \doi{10.1088/2041-8205/739/2/L63}
\bibAnnoteFile{torok2011}

\bibitem[{{Tousey}(1973)}]{tousey1973space}
{Tousey}, R. (1973).
\newblock {The solar corona}.
\newblock In \emph{Space Research Conference}, eds. M.~J. {Rycroft} and S.~K.
  {Runcorn} (Berlin: Akademic Verlag), vol.~2, 713--730
\bibAnnoteFile{tousey1973space}

\bibitem[{{Vourlidas} et~al.(2002){Vourlidas}, {Howard}, {Morrill}, and
  {Munz}}]{vourlidas2002analysis}
{Vourlidas}, A., {Howard}, R.~A., {Morrill}, J.~S., and {Munz}, S. (2002).
\newblock {Analysis of Lasco Observations of Streamer Blowout Events}.
\newblock In \emph{Solar-Terrestrial Magnetic Activity and Space Environment},
  eds. H.~{Wang} and R.~{Xu} (Boston: Pergamon), vol.~14, 201
\bibAnnoteFile{vourlidas2002analysis}

\bibitem[{{Vourlidas} and {Webb}(2018)}]{vourlidas2018streamer}
{Vourlidas}, A. and {Webb}, D.~F. (2018).
\newblock {Streamer-blowout Coronal Mass Ejections: Their Properties and
  Relation to the Coronal Magnetic Field Structure}.
\newblock \emph{The Astrophysical Journal} 861, 103.
\newblock \doi{10.3847/1538-4357/aaca3e}
\bibAnnoteFile{vourlidas2018streamer}

\bibitem[{{Wang} and {Sheeley}(1992)}]{wang1992potential}
{Wang}, Y.~M. and {Sheeley}, J., N.~R. (1992).
\newblock {On Potential Field Models of the Solar Corona}.
\newblock \emph{The Astrophysical Journal} 392, 310.
\newblock \doi{10.1086/171430}
\bibAnnoteFile{wang1992potential}

\bibitem[{{Whittlesey} et~al.(2020){Whittlesey}, {Larson}, {Kasper}, {Halekas},
  {Abatcha}, {Abiad} et~al.}]{whittlesey2020solar}
{Whittlesey}, P.~L., {Larson}, D.~E., {Kasper}, J.~C., {Halekas}, J.,
  {Abatcha}, M., {Abiad}, R., et~al. (2020).
\newblock {The Solar Probe ANalyzers{\textemdash}Electrons on the Parker Solar
  Probe}.
\newblock \emph{The Astrophysical Journal Supplement Series} 246, 74.
\newblock \doi{10.3847/1538-4365/ab7370}
\bibAnnoteFile{whittlesey2020solar}

\bibitem[{{Zheng} et~al.(2020){Zheng}, {Chen}, and {Wang}}]{Zheng2020}
{Zheng}, R., {Chen}, Y., and {Wang}, B. (2020).
\newblock {The Initiation of a Solar Streamer Blowout Coronal Mass Ejection
  Arising from the Streamer Flank}.
\newblock \emph{The Astrophysical Journal Letters} 897, L21.
\newblock \doi{10.3847/2041-8213/ab9ebd}
\bibAnnoteFile{Zheng2020}

\bibitem[{{Zhou} et~al.(2021){Zhou}, {Shen}, {Zhou}, {Tang}, {Duan}, and
  {Tan}}]{2021ApJ...923...45Z}
{Zhou}, C., {Shen}, Y., {Zhou}, X., {Tang}, Z., {Duan}, Y., and {Tan}, S.
  (2021).
\newblock {Sympathetic Filament Eruptions within a Fan-spine Magnetic System}.
\newblock \emph{The Astrophysical Journal} 923, 45.
\newblock \doi{10.3847/1538-4357/ac28a0}
\bibAnnoteFile{2021ApJ...923...45Z}

\end{thebibliography}

\end{document}